\pdfoutput=1
\documentclass[11pt]{article}
\usepackage[utf8]{inputenc}
\usepackage[T1]{fontenc}
\usepackage{lmodern}
\usepackage[margin=2.2cm]{geometry}
\usepackage{amsmath}
\usepackage{graphicx}
\usepackage{array}
\usepackage{booktabs}
\usepackage[table]{xcolor}
\usepackage{bm}
\usepackage{authblk}
\usepackage[round,authoryear]{natbib}
\usepackage[hidelinks]{hyperref}
\newcommand\blfootnote[1]{%
	\begingroup
	\renewcommand\thefootnote{}\footnote{#1}%
	\addtocounter{footnote}{-1}%
	\endgroup
}
\title{Joint Controls of Stem Water Storage, Root Conductance, and Rhizosphere Radial Flow on Hydraulic Redistribution}
\author[1]{Mahmoud Mbarak}
\author[1,2]{Jean-Christophe Domec}
\author[3,4]{Marc A. Hesse}
\author[5]{Sanna Sevanto}
\author[6]{Gabriel G. Katul}
\affil[1]{{\small Nicholas School of the Environment, Duke University, Durham, NC}}
\affil[2]{{\small Bordeaux Sciences Agro, Institute of Agricultural and Forest Sciences, Bordeaux, France}}
\affil[3]{{\small Department of Earth and Planetary Sciences, University of Texas, at Austin, Austin, TX}}
\affil[4]{{\small Oden Institute for Computational Engineering and Sciences, The University of Texas, at Austin, Austin, TX}}
\affil[5]{{\small Earth and Environmental Sciences Division, Los Alamos National Laboratory, Los Alamos, NM}}
\affil[6]{{\small Department of Civil, Construction, and Environmental Engineering, University of Alabama, Tuscaloosa, AL}}
\date{}

\begin{document}
\maketitle

\begin{abstract}
Hydraulic redistribution (HR) moves water from wet to dry soil directly through plant roots thereby 'short-circuiting' the slower movement of water across multiple soil layers.  The ecological significance of such water releases associated with HR is not in dispute as they tend to ameliorate and delay the onset of water stress in plants.  Several factors influencing HR have been identified, including soil-root water potential gradients, night-time transpiration, plant water storage, hydraulic conductance of the xylem and soil-root interface, and water transport within the rhizosphere. A comprehensive picture of their combined impact on HR and its magnitude relative to transpiration and root water uptake remains a subject of debate and active research.
To determine when a transpiring plant can supply water to the soil, a soil--root--xylem hydraulics model was developed to resolve radial flow in the rhizosphere and vertical flow across soil layers and a porous xylem, with root-collar water potential solved implicitly. Stomatal conductance was computed assuming a carbon-gain optimization constrained by soil water content. Storage capacity, axial xylem conductance, the timing of transpiration during the day, the root density profile, the rhizosphere gradient and root conductance were each changed over a 20-day  soil dry-down period.
The model shows that root water release to the soil occurred only when the storage relaxation time, $\tau$, set by the storage capacity and root and xylem conductances, was shorter than or comparable to the length of the nighttime period  of negligible transpiration, allowing sufficient xylem refilling. Further, water release from roots required a diurnal cycle of transpiration and a vertical soil-head contrast larger than the residual pull of the stem on the collar, and across the factorial ensemble fell by a median of 98\% (84--100\%) when the linear root profile was made uniform, and by a median of 99.8\% (83--100\%) when the rhizosphere gradient was resolved.  A Shapley-type decomposition of $\log_{10}$-transformed root water release over 450 full-factorial diurnal runs assigned 42\% of the variance to storage capacity, 32\% to root conductance, 12\% to the rhizosphere gradient, 8\% to the root profile and 7\% to axial xylem conductance, with interactions among factors accounting for 39\% of the total variance. The stem storage that buffers xylem tension by day is thus the same property that limits the redistribution process at night.

\vspace{0.5em}
\end{abstract}

\blfootnote{E-mail addresses: mahmoudmbarak95@gmail.com}

\section{Introduction}
\label{sec:intro}

The significance of hydraulic redistribution (HR), the passive movement of water directly through plant roots from wetter to drier soil, to the water balance of water-limited ecosystems is not in dispute \citep{caldwell1998,neumann2012,prieto2012}.
Water transfer through roots from moist to dry soil was shown in pot experiments in the 1930s and 1940s \citep{breazeale1930,breazeale1934,volk1947}, and the process was termed 'hydraulic lift' in a field stand of \textit{Artemisia tridentata} \citep{richards1987}.
HR has since been measured as upward, downward and lateral flow \citep{burgess1998,hultine2004,bleby2010} in deserts, savannas, woodlands and pine forests \citep{dawson1993,scholz2002,brooks2002,meinzer2004,warren2007,domec2010}.
Its reported average magnitude varies by nearly two orders of magnitude, from 0.04 to 1.3~mm~d$^{-1}$ in field studies and from 0.1 to 3.23~mm~d$^{-1}$ in models \citep{neumann2012}.
A global synthesis of 47 sites gave a weighted mean of 0.249~mm~d$^{-1}$ and a mean contribution to daily transpiration of 27.4\%, ranging from 3 to 79\% \citep{yang2022}.
Lifted water was estimated at about one third of daily evapotranspiration (ET) for \textit{A.~tridentata} and at 14\% of daily ET for a desert subshrub \citep{caldwell1998}.
HR replaced about 28\% and 35\% of the daily depletion of the upper 2~m of soil in Douglas-fir and ponderosa pine stands \citep{brooks2002}, and it amounted to about 22\% of daily transpiration in a semi-arid woodland rooted to 20~m \citep{bleby2010}.
In old-growth conifer forests it accounted for only 3--9\% of seasonal site water depletion, at 0.16--0.30~mm~d$^{-1}$ at its peak \citep{warren2007}, and across six forest and savanna sites it stayed between 0 and 0.4~mm~d$^{-1}$ in the upper 60~cm of soil \citep{meinzer2004}.
The consequences of HR have been discussed from a local scale (roots) to the climate system.
Hydraulic redistribution can mitigate embolism in shallow roots during drought \citep{domec2004}. In two conifer stands, HR was estimated to conserve upper-soil water equivalent to an additional 16--21 days of water use following a 60-day drought \citep{brooks2002}. A hydraulic model constrained by field observations estimated that HR increased tree transpiration by 30--50\% in a drought-stressed pine plantation \citep{domec2010}, while an atmospheric general circulation model estimated an increase of approximately 40\% in dry-season transpiration over the Amazon \citep{lee2005}.  Through its effect on the diurnal course of surface fluxes, root water redistribution also enters the growth of the convective boundary layer \citep{siqueira2009}, and it becomes important for sustaining ET over the course of a drought \citep{amenu2008,katul2012}.  However, the conditions under which plants release water into dry soil at night, and the magnitude of this flux relative to daytime transpiration, remain incompletely understood.

Field studies identify the factors affecting the night-time water potential gradient that drives HR \citep{caldwell1998,prieto2012} (Table~\ref{tab:field}).
The first is the spatial heterogeneity of soil water potential (vertically and across the rhizosphere) spanned by the root system. HR can increase as the upper soil dries, provided that wetter soil layers remain accessible and root hydraulic conductivity is maintained \citep{meinzer2004}. HR also depends on the conductance of the root pathway, and it declined late in the season as the modeled pathway conductance fell with root cavitation \citep{warren2007}.
The second factor is night-time transpiration.
Suppressing it increased HR in two of three species \citep{howard2009}, and in the field reverse root flow or HR declined as night-time evaporative demand or nocturnal transpiration rose \citep{hultine2003,domec2010,yu2018}.
The third, and perhaps a less appreciated factor, is the refilling of plant water storage.  Stored water supplies part of daily transpiration \citep{scholz2007}, and capacitance buffers the diurnal fluctuations of water potential in stems and roots \citep{meinzer2009,domec2006,scholz2007}.
Overnight recharge is usually incomplete in species with more storage \citep{scholz2007}. In those cases, HR was more strongly related to stem refilling than to nocturnal transpiration \citep{yu2018}, and storage withdrawal and refilling were proposed to have prevented HR in a desert phreatophyte \citep{hultine2003}.
In a soil--plant model of beech and spruce, a larger capacitance raised night-time root water uptake, because more storage emptied by day had to be refilled \citep{martinetti2026}.
These observations establish the factors, but nocturnal sap flux is difficult to partition into transpiration and refilling \citep{wang2026}, and no field study can change storage while holding soil water, roots and demand fixed.  Such inquiries do necessitate mathematical models that can hold key state variables constant while altering others.
Models of HR differ most in how they represent the plant at night (Table~\ref{tab:models}).
Models without a root water potential move water directly between soil layers and need an on/off term to restrict HR in the night \citep{ryel2002,montaldo2022}.
Models that route uptake and release through a root water potential but carry no plant storage let that potential follow transpiration without lag \citep{mendel2002,siqueira2008,volpe2013,manoli2014}.
As transpiration declines at dusk, root xylem water potential can become intermediate between those of wetter and drier rooted soil layers, allowing water to move through the roots from wetter to drier soil \citep{siqueira2008}.
Both model classes predict HR fluxes equivalent to a substantial fraction of transpiration (Table~\ref{tab:models}). However, their predictions differ when plant water storage is included. 
In a well-watered season, a single whole-plant capacitance in a one-dimensional soil model limited HR to occasional major rainfall events, when transpiration was set to zero \citep{vogel2017}.
In multilayer and porous-media models, HR persisted with storage \citep{domec2012,huang2017,manoli2017} and was reduced by a larger capacitance \citep{huang2017} and by nocturnal transpiration \citep{domec2012,huang2017}, which leave a residual gradient along the plant that competes with the soil at night \citep{huang2017}.

What is missing is (i) a comprehensive condition for root release that holds at night, when uptake also refills plant storage and (ii) the relative contribution of each control to the variance of HR, measured within a single model at fixed soil and daily demand.  Model outcomes range from HR only during major rainfall events \citep{vogel2017} to HR that amounts to most of tree transpiration \citep{montaldo2022}.
The competition with storage has been described as a sink strength that changes the amount of HR \citep{huang2017} or as a contrast between zero and one fitted capacitance \citep{vogel2017}, and the models with storage did not report how fast it relaxes relative to the night \citep{huang2017,manoli2017}.
On the soil side, the effect of radial gradients in rhizosphere water potential has not been isolated within a single model by comparing simulations with and without these gradients while holding soil properties, root architecture, and atmospheric demand constant. 

Boosted regression trees have related HR to transpiration, root length, water-table depth and a humidity index across 35 field and 12 modelling estimates \citep{yang2022}. These variables covary across sites, and neither plant water storage nor root hydraulic conductance was included among the predictors.  

The necessary condition for root release of \citet{couvreur2012} compares the local soil head with transpiration, which at night differs from the uptake that refills xylem storage.
Deriving the missing night-time condition and measuring the contribution of each control requires a single model that resolves the diurnal cycle, the radial rhizosphere gradient, the root profile, the axial xylem conductance, the root conductance, and storage. Such a model enables each control to be changed in isolation and in combination at fixed soil type and for prescribed repeating meteorological drivers so that release can be attributed to them.  To address (i) and (ii), a coupled soil--root--xylem model was developed and used first in single configurations and then as an ensemble.
The analysis uses the storage relaxation time $\tau = C/A_s$, where $C$ is the xylem storage capacity and $A_s$ is the effective conductance between the soil and the stored water, set by the root and xylem-column conductances in series.

The model developed here seeks to accomplish the following:
\begin{enumerate}
  \item Couples a two-scale soil model with radial flow in the rhizosphere and vertical Richards flow \citep{siqueira2008} to a porous-media xylem column with pressure-dependent storage \citep{bohrer2005,silva2022}, with the root-collar potential solved implicitly and stomatal conductance inferred from a carbon-gain optimization \citep{katul2010}.
  \item Reads each run through the condition for root release, which compares the soil-head contrast seen by the roots with the pull of net uptake on the collar.
  \item Runs a full-factorial ensemble of 900 runs over root conductance, root profile, storage capacity, axial conductance, the rhizosphere gradient and the timing of transpiration.
  \item Attributes release in the ensemble to the controls with an exact Sobol--Shapley decomposition, and separates the share each control holds alone from the share it holds through interactions.
  \item Compares the release of the ensemble with reported HR in the literature to assess whether overall trends are consistent with presumed causes.
\end{enumerate}

Simulations use a reference configuration and vary one control at a time: the temporal pattern of transpiration (a 12-h daytime period versus a constant rate over 24 h, with the same daily total), the radial gradient in rhizosphere water potential, the vertical root distribution, axial xylem conductance, root hydraulic conductance, and plant water storage capacity. To restrict the scope and compare across runs, the analysis fixes the soil type, the root-zone depth, the meteorological forcing, and the duration of drydown.  Those factors have been explored in other studies \citep{siqueira2008,siqueira2009,huang2017}.

\begin{table}[p]
\centering
\caption{Hydraulic redistribution (HR) in field and greenhouse studies: reported magnitude and its normalisation, and what each study reports on night-time transpiration ($E_n$). Shares are given in the normalisation each study used and are not interchangeable. n.r., not reported.}
\label{tab:field}
\vspace{4pt}
\footnotesize
\setlength{\tabcolsep}{4pt}
\renewcommand{\arraystretch}{1.3}
\rowcolors{2}{black!5}{white}
\begin{tabular}{@{}>{\raggedright\arraybackslash}p{2.7cm}>{\raggedright\arraybackslash}p{3.5cm}>{\raggedright\arraybackslash}p{6.0cm}>{\raggedright\arraybackslash}p{4.0cm}@{}}
\toprule
\rowcolor{white}\textbf{Study} & \textbf{System} & \textbf{HR magnitude and share} & \textbf{$\boldsymbol{E_n}$} \\
\midrule
\citet{richards1987}; \citet{caldwell1998} & \textit{A.~tridentata}, Utah & $\approx 1$~L~m$^{-2}$ per night, roughly one third of daily ET, as summarized by \citet{caldwell1998} & night-time illumination suppressed the nocturnal rise of $\psi_s$ \citep{caldwell1998} \\
\citet{brooks2002} & 20-yr Douglas-fir and old-growth ponderosa pine & $\approx 28$\% (late August) and $\approx 35$\% (July--August) of daily water use from the upper 2~m replaced by HR & n.r. \\
\citet{meinzer2004} & 3 conifer forests (USA), 3 savanna sites (Brazil) & 0--0.4~mm~d$^{-1}$ into the upper 60~cm; overnight recharge rose sharply below $\psi_s \approx -0.4$~MPa, to 80--90\% of the previous day's use & n.r. \\
\citet{warren2007} & old-growth Douglas-fir and ponderosa pine & 0--0.5~mm~d$^{-1}$ within sites, peak 0.16--0.30~mm~d$^{-1}$ & night VPD among tested correlates of HR \\
\citet{domec2010} & 18-yr loblolly pine plantation, drought & peak 1.05~mm~d$^{-1}$; nightly backflow $\approx 15$\% of daily upward flow in the 2 of 3 monitored lateral roots with reverse flow & negative relation of HR to nocturnal $T$ (data not shown) \\
\citet{bleby2010} & semi-arid woodland, roots to 20~m & up to $\approx 22$\% of daily $T$ (HR volume over 1 month) & night flow from capacitance or $E_n$ separated from HR by its time course \\
\citet{scholz2002} & 8 Cerrado species & reverse root flow in 4 species, at night in some and day and night in others & n.r. \\
\citet{hultine2003} & 3 desert phreatophytes, one tree each & reverse flow up to 120~g~h$^{-1}$ (\textit{Fraxinus}) and 18~g~h$^{-1}$ (\textit{Juglans}); none in \textit{Celtis}, where storage withdrawal and refilling were proposed to have prevented HR & \textit{Fraxinus} reverse flow inversely correlated with night VPD; no relation in \textit{Juglans} \\
\citet{howard2009} & 3 species, split-root pots, greenhouse & HR as the night-time rise of $\psi_s$ in the dry compartment & bagging raised nightly HR by 73\% and 33\% in two species, no effect in the third \\
\citet{yu2018} & \textit{Populus euphratica}, riparian & HR more strongly related to stem refilling ($R^2 = 0.41$) than to $E_n$ ($R^2 = 0.14$) & $\approx 20$\% of nocturnal sap flow, refilling $\approx 80$\% \\
\bottomrule
\end{tabular}
\end{table}

\begin{table}[p]
\centering
\caption{Hydraulic redistribution (HR) in models: reported magnitude and its normalisation, and how each model represents plant water storage, night-time transpiration ($E_n$) and the rhizosphere. Day/night gating is noted where it is imposed. Shares are given in the normalisation each study used and are not interchangeable. n.r., not reported.}
\label{tab:models}
\vspace{4pt}
\footnotesize
\setlength{\tabcolsep}{4pt}
\renewcommand{\arraystretch}{1.3}
\rowcolors{2}{black!5}{white}
\begin{tabular}{@{}>{\raggedright\arraybackslash}p{2.1cm}>{\raggedright\arraybackslash}p{2.9cm}>{\raggedright\arraybackslash}p{3.4cm}>{\raggedright\arraybackslash}p{2.4cm}>{\raggedright\arraybackslash}p{2.3cm}>{\raggedright\arraybackslash}p{2.7cm}@{}}
\toprule
\rowcolor{white}\textbf{Study} & \textbf{System} & \textbf{HR magnitude and share} & \textbf{Plant storage} & \textbf{$\boldsymbol{E_n}$} & \textbf{Rhizosphere} \\
\midrule
\citet{ryel2002} & 1-D soil layers, \textit{A.~tridentata}; HR at night only & 5.98~cm over a 100-d summer with one rain event, 19\% of transpiration & none & none & vertical soil, root conductance \\
\citet{mendel2002} & 2-D soil and root-water-potential fields, sugar maple setting & 172~L~d$^{-1}$ per tree, above the field estimate of $102 \pm 54$~L~d$^{-1}$ & none & $T$ a Gaussian daily curve (peak 14:00, s.d.\ 3~h) & root radial conductance \\
\citet{siqueira2008} & radial + vertical Richards, generic stand, 1-m roots & maximum 1.14 (high $K_r$) and 0.80 (low $K_r$)~mm~d$^{-1}$, 37\% and 25\% of potential $T$ & none ($T$ = sap flow) & residual stomatal conductance $g_0$ & radial gradient resolved \\
\citet{volpe2013} & multilayer soil--plant--atmosphere, grassland, simulated drydown & $\approx 12$\% of diurnal $T$ supplied directly by root HR (roots concentrated near the surface) & none (assumed negligible) & n.r. & vertical soil and root conductance \\
\citet{manoli2014} & 3-D soil + trunk xylem, loblolly pine & peak 0.5 (2007) and 1.1 (2008)~mm~d$^{-1}$, 10--20\% of daily $T$ & none (assumed) & $g_{s,n}$ = 18 mmol m$^{-2}$ s$^{-1}$ & soil conductance (radial path to the nearest rootlet) and root-membrane conductance in series to one root-xylem potential \\
\citet{domec2012} & multilayer soil--plant--atmosphere model, three loblolly pine plantations & root leakage 7--45\% of growing-season root uptake, current climate & one capacity per species & cut leakage by 25--36\% & vertical soil, root conductance \citep{braud1995} \\
\citet{huang2017} & porous-media soil--plant, eight scenarios & mean 6.3--16.7\% of daily $T$ & stem ($C_p$) and leaf ($C_l$, Eq.~10); larger $C_p$ cut HR by $>22$\% & removing it raised HR $\approx 10$\% & as in \citet{manoli2014} \\
\citet{manoli2017} & 3-D soil + porous-medium stem, loblolly pine & 10--30\% of $T$ & resolved along stem & cuticular conductance allows night losses & as in \citet{manoli2014} \\
\citet{vogel2017} & 1-D soil, Norway spruce, well-watered season & nightly with $C = 0$; with fitted $C$, only occasionally during major rainfall events & one whole-plant capacitance & none & rhizosphere radius, resistance in series; rhizosphere not solved \\
\citet{montaldo2022} & buckets (surface soil, rhizosphere, fractured rock), wild olive; HR at night only (as in Ryel) & up to 0.9--1.0~mm~d$^{-1}$; 80--85\% of tree $T$ and 60--65\% of ET in the driest summers & none & n.r. & rhizosphere bucket \\
\bottomrule
\end{tabular}
\end{table}

\clearpage

\section{Methods}

\subsection{Model structure}
\begin{figure}
    \centering
    \includegraphics[width=\textwidth]{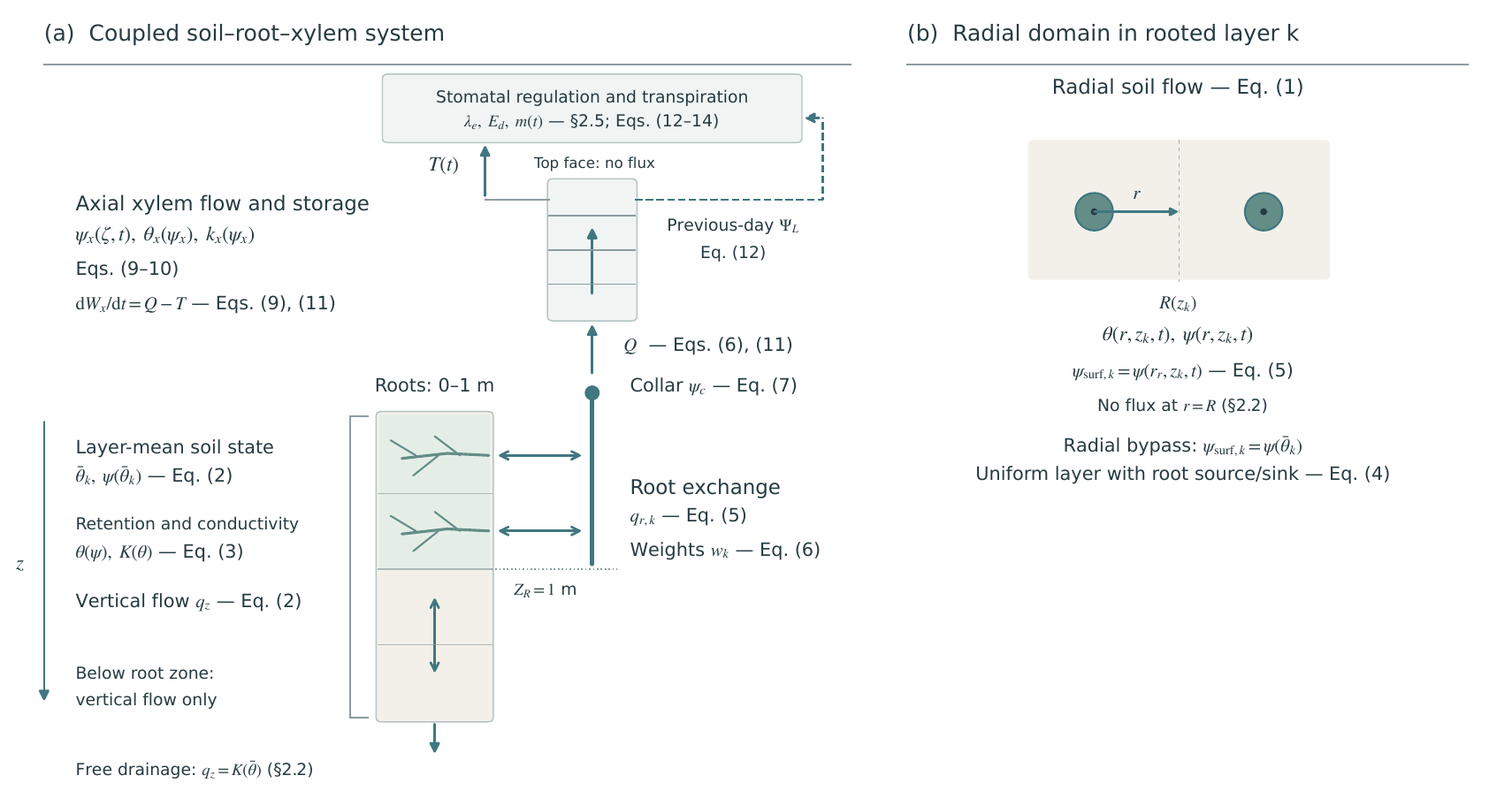}
    \caption{Model Structure. (a) Soil, roots, and xylem column coupled through the collar potential $\psi_c$, with the equations that govern each part. (b) Radial domain in rooted layer $k$, resolved (radial ON) or bypassed (radial OFF).}
    \label{fig:schematic}
\end{figure}

The numerical model assumes that water losses from leaves to sustain photosynthesis are first drawn from the upper layers of the xylem.  The declining water potential in these upper layers 'pulls' water from the lower xylem layers. The depleted xylem storage is replenished by soil water. This replenishment occurs by adjusting  water potential at the root interface so as to draw (or release) water from multiple soil layers. Mathematically, the model considers two-scale (radial and vertical) soil water transport \citep{siqueira2008} coupled to a porous-media xylem column with storage that depends on xylem water potential \citep{chuang2006,bohrer2005} through one root-collar potential (Fig.~\ref{fig:schematic}a).  Soil and roots exchange water across the root membrane in every rooted layer, and the collar potential is the unknown that matches this exchange to the flow entering the xylem.  Comparable couplings of a resolved soil to a porous-medium stem were used in prior studies \citep{manoli2017,huang2017}.
Stomatal conductance follows a carbon-gain optimization \citep{katul2010,manzoni2011} with its marginal water use efficiency adjusted daily to account for soil drying. The resulting daily transpiration is delivered in time by a prescribed mask.  Fluxes and storages are per unit ground area.  Soil and collar potentials are expressed in meters of water and the xylem water potential in pascals, and both are reported in MPa (1~MPa $\approx 102$~m).
The key model parameters are listed in Table~\ref{tab:params}.

Each simulation is an idealized 20-day rain-free dry-down driven by constant synthetic meteorological forcing (vapor pressure deficit $D$, CO$_2$ concentration $c_a$, photon flux density $Q_p$, and leaf temperature $T_L$). Before coupling to the soil, the xylem column is initialized through a 5-day wet spin-up (zero lower-boundary tension); at day 1 it is coupled to a loam column initialized at a uniform water content $\theta_0 = 0.4\,\theta_{sat} = 0.172$~m$^3$~m$^{-3}$. Days 1--2 are excluded as start-up transients, and all diagnostics are evaluated over days 3--20 (nights 3--19).
    
\subsection{Soil water flow}

Soil water flow follows the homogenization of \citet{siqueira2008}, which separates radial flow toward the roots, on scales of millimetres, from vertical flow through the profile, on scales of 10-100 cm.  Around the roots of the layer at depth $z$ (positive downward), the water content $\theta(r,z,t)$ obeys
\begin{equation}
\frac{\partial\theta}{\partial t}=\frac{1}{r}\frac{\partial}{\partial r}\left[rK(\theta)\frac{\partial\psi}{\partial r}\right]-\frac{\partial q_z}{\partial z},\qquad r_r\le r\le R(z),
\label{eq:radial}
\end{equation}
where $r$ is the distance from the root axis, $\psi$ is the matric potential, $K$ is the hydraulic conductivity, $q_z$ is the vertical Darcy flux, $r_r$ is the root radius and $R(z)=[\pi\lambda(z)]^{-1/2}$ is half the distance between roots of length density $\lambda$ \citep[Eq.~4a]{siqueira2008}.  That is, the root density profile dictates the separation distances between rootlets in a given soil layer at depth $z$.
The vertical flux is computed from the layer-mean water content \citep[Eqs.~4b and 5]{siqueira2008},
\begin{equation}
q_z=-K(\bar\theta)\left[\frac{\partial\psi(\bar\theta)}{\partial z}-1\right],\qquad \bar\theta(z,t)=\frac{2}{R^2-r_r^2}\int_{r_r}^{R}\theta\,r\,\mathrm{d}r .
\label{eq:vertical}
\end{equation}
The flux at the outer boundary, $r=R$, is zero by symmetry, and at the root surface, $r=r_r$, we apply the root exchange parameterization discussed in Section~\ref{sec:roots}.  For simplicity, the soil surface carries no flux, so neither rainfall nor soil evaporation enters the runs.  The base of the soil column drains freely under a unit head gradient (i.e. gravitational drainage), $q_z=K(\bar\theta)$, and below the root zone only vertical flow occurs.
Retention $S_e$  and conductivity $K$ follow \citet{vangenuchten1980} with the pore-connectivity model of \citet{mualem1976},
\begin{equation}
S_e=\frac{\theta-\theta_{res}}{\theta_{sat}-\theta_{res}}=\left[1+(\alpha|\psi|)^{n}\right]^{-m},\qquad K=K_sS_e^{1/2}\left[1-\left(1-S_e^{1/m}\right)^{m}\right]^{2},\qquad m=1-\frac{1}{n},
\label{eq:vg}
\end{equation}
with loam parameters of \citet{carsel1988} selected as reference for $K_s$, $n$, $\alpha$, $\theta_{res}$, and $\theta_{sat}$.  Soil water content is bounded below by its value at a minimum potential $\psi_{min}$.

The rhizosphere gradient was bypassed (radial OFF) by reducing each rooted layer to one uniform volume, so that the potential at the root surface equals $\psi(\bar\theta)$.  Eq.~(\ref{eq:radial}) reduces to the conventional vertical Richards equation with a root water uptake sink $S$ given by
\begin{equation}
\frac{\partial\bar\theta}{\partial t}=-\frac{\partial q_z}{\partial z}-S, \qquad S=-2\pi r_r\lambda~ q_r.
\label{eq:Richard_vertical}
\end{equation}
With the radial gradients resolved (radial ON), Eq.~(\ref{eq:radial}) was solved on a fine radial grid in every rooted layer (Fig.~\ref{fig:schematic}b).

\subsection{Root water exchange and the collar potential}
\label{sec:roots}

Water crosses the root surface at the rate \citep[Eq.~6]{siqueira2008}
\begin{equation}
q_r(z,t)=K_r\left[\psi_c(t)+z-\psi_{surf}(z,t)\right]
\label{eq:qr}
\end{equation}
per unit root surface area, positive from root to soil.  Here $\psi_c$ is termed the root-collar potential referenced to the soil surface, $\psi_{surf}=\psi(r_r,z,t)$ is the matric potential at the root surface, and $K_r$ is the root membrane permeability.  Pressure losses inside the roots and osmotic transport are neglected, so every root carries the collar potential plus its hydrostatic offset $z$ \citep{siqueira2008,katul2010hr}.
The $K_r$ was treated as constant throughout the model but the model allows for adjustments due to root-soil separation.  Three types of root length density were considered: decreasing linearly from the surface to the rooting depth $Z_R$, $\lambda=2\bar\lambda(1-z/Z_R)$ (profile B0), uniform, $\lambda=\bar\lambda$ (profile B0R) \citep[Table~1 and Fig.~3]{siqueira2008}, or bimodal (profile B0M), with the same mean $\bar\lambda$ in all three.  The half-distance $R$ thus increases from 6.3~mm at the surface to 63~mm at the base of the root zone for B0 and equals 8.9~mm for B0R. A bimodal profile that places 10\% of the root length uniformly and 90\% in two Gaussian modes of equal weight at 0.15 and 0.85~m depth, each 0.08~m wide is also explored so that $\lambda=\bar\lambda Z_R\left[0.1/Z_R+0.9\left(0.5\,\mathcal{N}(0.15,0.08)+0.5\,\mathcal{N}(0.85,0.08)\right)\right]$ with each mode normalised over the root zone. It has the same total root length and mean root depth (0.5~m) as B0R, and $R$ ranges from 5.7 to 28.2~mm.
Summed over the $N$ rooted layers of thickness $\Delta z$, the net root uptake per unit ground area is
\begin{equation}
Q(t)=-\sum_{k=1}^{N}w_kq_{r,k},\qquad w_k=2\pi r_r\lambda_k\Delta z. 
\label{eq:Q}
\end{equation}
Combining Eqs.~(\ref{eq:qr}) and (\ref{eq:Q}) gives the collar potential,
\begin{equation}
\psi_c=H-\frac{Q}{A_r},\qquad H=\frac{\sum_kw_kh_k}{\sum_kw_k},\qquad h_k=\psi_{surf,k}-z_k,\qquad A_r=K_r\sum_kw_k 
\label{eq:collar}
\end{equation}
where $h_k$ is the total head (matric potential minus depth) of the soil at the root surface of layer $k$ representing the gravitational potential and $A_r$ is the conductance of the root system.
The $H$ is not a mean soil water content.  It is the root-length-weighted mean of $h_k$, with weights $w_k\propto\lambda_k$, so densely rooted layers count more.
Eq.~(\ref{eq:collar}) is Eq.~8 of \citet{couvreur2012}, with the net uptake $Q$ in place of transpiration.  It follows that layer $k$ releases water ($q_{r,k}>0$) when
\begin{equation}
H-h_k>\frac{Q}{A_r},
\label{eq:release}
\end{equation}
that is, when the soil-head contrast between the layer and the root-weighted mean exceeds the depression of the collar below $H$.

\subsection{Xylem water flow and storage}
\label{sec:xylem}

Water flow in the xylem follows a porous-media like formulation of \citet{chuang2006} and \citet{bohrer2005}, reduced to one vertical column of height $L$ whose cross-section is the unit ground area,
\begin{equation}
\frac{\partial\theta_x}{\partial t}=\frac{\partial}{\partial\zeta}\left[k_x(\psi_x)\left(\frac{\partial\psi_x}{\partial\zeta}+\rho_wg\right)\right]\pm S_x,\qquad 0<\zeta<L,
\label{eq:xylem}
\end{equation}
where $\zeta$ is the height above the collar, $\psi_x$ is the xylem water potential (Pa), $\theta_x$ is the 'mobile' xylem water content (kg~m$^{-3}$), $k_x$ is the xylem conductivity (kg~m$^{-1}$~s$^{-1}$~Pa$^{-1}$), $S_x$ is a source or sink of water and $\rho_w g$ is the specific weight of water formed by the density $\rho_w$ and the gravitational acceleration $g$.
Eq.~(\ref{eq:xylem}) is written in mixed form, with storage in $\theta_x$ and flux in $\psi_x$, so that the discretized equation conserves water mass \citep{aziz1979petroleum}.
\citet{bohrer2005} wrote it in the capacitance form, $C(\psi_x)\,\partial\psi_x/\partial t$, and \citet{chuang2006} in terms of $\theta_x$ alone.
Conductivity and storage follow \citet{bohrer2005}
\begin{equation}
k_x=f_Kk_{max}\exp\left[-\left(\frac{-\psi_x}{c_1}\right)^{c_2}\right],\qquad \theta_x=f_C\,\theta_{x,sat}\left(\frac{P_0-\psi_x}{P_0}\right)^{-p},
\label{eq:fetch}
\end{equation}
where the factors $f_K$ and $f_C$, added here, scale conductivity and storage independently.
The storage capacity per unit ground area, $C=\rho_w^{-1}\int_0^L(\mathrm{d}\theta_x/\mathrm{d}\psi_x)\,\mathrm{d}\zeta$, is $192f_C$~mm~MPa$^{-1}$ at $\psi_x=0$.  

We point out that an axial porous media approximation for xylem water movement, which accounts for wall friction, ignores head losses due to lateral transport across vessels or tracheids separated by membranes.  However,  those lateral head losses, commonly derived from a Sampson flow approximation, remain proportional to their axial counterpart in realistic xylem networks \cite{mrad2018network}.  That is, xylem networks have evolved their configuration so that their axial and lateral head losses are coordinated and remain commensurate with each other. Resolving one of them (axial here) informs the head loss throughout the entire xylem network as assumed here. Thus, the hydraulic properties of the xylem selected here are presumed to be 'effective' so as to accommodate both axial and radial frictional losses.  Such effective representation is only permissible of the axial (wall friction) and radial (membrane) head losses are comparable.  At the collar, the xylem water potential equals the collar potential, $\psi_x(0,t)=\rho_wg\psi_c(t)$, so that water potential is continuous between roots and xylem.  Transpiration $T(t)$ is withdrawn from the uppermost cell of the xylem column by leaves, and the top face carries no other flux.  At all times $\psi_c$ satisfies
\begin{equation}
-k_x\left(\frac{\partial\psi_x}{\partial\zeta}+\rho_wg\right)_{\zeta=0}=\rho_wQ ,
\label{eq:coupling}
\end{equation}
so that the flow entering the xylem equals the net root uptake of Eq.~(\ref{eq:Q}).
Without storage $Q$ would equal $T$ at every instant, as assumed by \citet{siqueira2008}.  With storage, $Q-T$ is the overall rate of change of xylem water content and dictates the emptying/refilling rates of the xylem during the course of the day.  However, xylem emptying and refilling are not uniform along the xylem.

\subsection{Stomatal conductance and transpiration}

Stomatal conductance to CO$_2$, $g_l$ (mol~m$^{-2}$~s$^{-1}$ per unit leaf area), maximizes the net carbon gain $A(g_l)-\lambda_e a g_l D$ \citep{katul2009,katul2010,manzoni2011}.
Here $A$ is the net assimilation rate, $a$ is the ratio of the diffusivities of water vapour and CO$_2$, $D$ is the vapour pressure deficit as a mole fraction and $\lambda_e$ is the marginal water use efficiency that dictates the cost of losing water molecules to the carbon economy of the plant.
Biochemical parameters and their temperature functions follow \citet{campbell1998}, with the carboxylation capacity $V_{c,max}$ of \citet{katul2010}.
The marginal water use efficiency is assumed to gradually increase as the plant dries,
\begin{equation}
\lambda_e=\lambda_1\exp\left[-\beta\left(\Psi_L-\Psi_1\right)\right],
\label{eq:lambda}
\end{equation}
where $\Psi_L$ is the total water potential at the top of the column at the end of the preceding day.  This choice ensures that $\lambda_e$ is 'quasi-constant' at the time scale of stomatal opening and closure (30 min) but increases from day-to-day as the dry down progresses.

Coupling Fickian diffusion $A = g_l(c_a - c_i)$ with the Farquhar C$_3$ biochemical demand $A = a_1(c_i - \Gamma^*)/(a_2 + c_i) - R_d$ (where $c_i$ is intercellular CO$_2$, $\Gamma^*$ is the CO$_2$ compensation point, $R_d = 0.01\,V_{c,max}$ is dark respiration, and $(a_1,a_2)$ are the temperature-adjusted parameters at leaf temperature $T_L = 20\,^\circ\mathrm{C}$) and setting $\partial(A - \lambda_e a g_l D)/\partial g_l = 0$ yields the closed-form optimal conductance, extending \citet[Eq.~8]{katul2010} to retain dark respiration, 
\begin{equation}
    \begin{aligned}
    g_l &= -\frac{a_1(a_2 - c_a + 2\Gamma^*) + (a_2 + c_a)R_d}{(a_2 + c_a)^2} \\
        &\quad + \frac{a_2 + c_a - 2aD\lambda_e}{(a_2 + c_a)^2}\sqrt{\frac{a_1(a_2 + \Gamma^*)\big[a_1(c_a - \Gamma^*) - (a_2 + c_a)R_d\big]}{aD\lambda_e\,(a_2 + c_a -           
        aD\lambda_e)}},
    \end{aligned}
    \label{eq:gopt}
\end{equation}
evaluated on both the Rubisco-limited and light-limited ($Q_p$-dependent) branches, selecting the branch that yields the lower net assimilation $A = g_l(c_a - c_i)$.
Conductance $g_l$ is computed once per day from Eq.~(\ref{eq:gopt}) (with $\lambda_1$, $\Psi_1$, and $\beta$ in Table~\ref{tab:params}) and held constant within the day.

The daily transpiration per unit ground area is
\begin{equation}
E_d=v\,a g_l D,\qquad v=\frac{\mathrm{LAI}\,t_dM_w}{\rho_w},
\label{eq:Ed}
\end{equation}
where $v$ converts the leaf-level flux $a g_l D$ to a daily depth per unit ground area, with leaf area index $\mathrm{LAI}$, daylight period $t_d$ and $M_w$ the molar mass of water.
Diurnal variation therefore enters only through the timing of the flux, $T(t)=E_d\,m(t)/(1~\mathrm{d})$, where the mask $m(t)$ has a daily mean of one.
With diurnal forcing (diurnal ON), the mask is
\begin{equation}
m(t)\propto\left[1+\tanh\frac{t-t_{dawn}}{w}\right]\left[1+\tanh\frac{t_{dusk}-t}{w}\right],
\label{eq:mask}
\end{equation}
with $w=28.8$~min and the day centred on noon.
The night lasts $t_{night}=12$~h (dawn at 06:00, dusk at 18:00), and the mask then places 2.8\% of the daily total outside 06:00--18:00.
Without diurnal forcing (diurnal OFF), $m=1$ and the same daily total is delivered at a constant rate.
Because $g_l$ responds to $\Psi_L$, the daily totals of paired runs can differ through this feedback.

\subsection{Diagnostics}
\label{sec:diag}

\paragraph{Storage relaxation time.}
The storage relaxation time (or e-folding time) can be defined as $\tau=C/A_s$, where $C$ is the storage capacity of Section~\ref{sec:xylem} averaged over days 3--20 and $A_s$ is the effective conductance between the soil and the stored water.
With $T=0$ (e.g. night-time) and $H$ quasi-steady, the storage balance $C\,\mathrm{d}\bar\Phi/\mathrm{d}t=Q=A_s(H-\bar\Phi)$, where $\bar\Phi$ is the mean potential of the stored water, is a linear first-order ODE, and $\tau=C/A_s$ is the characteristic time scale of the decay of $Q$.

For a store distributed uniformly along the column, the root system and half of the column act in series, $A_s\approx(1/A_r+1/2K_x)^{-1}$, where $K_x=f_Kk_{max}/L$ is the axial conductance of the column.
For the runs with the rhizosphere gradient resolved, Section~\ref{sec:rhizo} also gives $\tau$ with $A_r$ replaced by the conductance of the rhizosphere and the root membrane in series, summed over the rooted layers and averaged over each night.
Elsewhere $\tau$ omits the rhizosphere, so a run with the gradient resolved has the $\tau$ of its partner with the gradient bypassed.

\paragraph{Attribution of release to the controls.}
Release $R$ is the positive part of the layer flux $w_kq_{r,k}$, summed over the rooted layers and over the time steps of days 3--20, $R=\sum_n\sum_k\max(w_kq_{r,k},0)\,\Delta t$.
It was decomposed on a balanced full-factorial run set in which every combination of the tested levels was run once.
The set crosses root conductance ($K_r=10^{-8}$, $10^{-9}$ and $10^{-10}$~s$^{-1}$), root profile (B0, B0R and B0M), storage ($f_C=0.01$, 0.1, 0.2, 0.5 and 1), axial conductivity ($f_K=0.1$, 0.5, 1, 2 and 10), the rhizosphere gradient (radial OFF and ON) and the timing of transpiration (diurnal ON and OFF), 900 runs in total. Within this run set, the reference run used root profile B0, $f_C=0.1$ (written $\times0.1$ in the figures), $f_K=1$, $K_r=10^{-8}$~s$^{-1}$, a 12-h night, the rhizosphere gradient bypassed (radial OFF) and diurnal transpiration (diurnal ON). No diurnal-OFF run released water after the start-up transient (Section~\ref{sec:diurnal}), so only the 450 diurnal-ON runs were decomposed, over the five remaining controls.
The model is deterministic, so the decomposition is exact and has no error term.
For a set $u$ of controls, the closed variance $V_u$ is the variance over runs of the mean release grouped by the levels of $u$, $V_u=\mathrm{Var}(E[y\mid X_u])$, with $y=\log_{10}(R+0.001~\mathrm{mm})$, which keeps the non-releasing runs in the design.
The Sobol terms \citep{sobol2001,owen2014} follow by inclusion--exclusion, $S_u=V^{-1}\sum_{v\subseteq u}(-1)^{|u|-|v|}V_v$, where $V$ is the total variance, and they sum to 1 over the 31 non-empty sets of the five controls.
The first-order index of control $i$ is $S_i=V_{\{i\}}/V$, its total index $S_{Ti}$ is the sum of all terms that contain $i$, and $S_{Ti}-S_i$ is its share held through interactions.
The Shapley share of control $i$ splits each term equally among its controls, $\mathrm{Sh}_i=\sum_{u\ni i}S_u/|u|$ \citep{owen2014}, so that $S_i\le\mathrm{Sh}_i\le S_{Ti}$ and $\sum_i\mathrm{Sh}_i=1$.
Each tested level carries equal weight, so the shares describe the tested ranges and not a population of plants.
The levels were therefore chosen to make the tested ranges wide: root conductance and storage capacity each span two orders of magnitude ($C$ from 1.9 to 192~mm~MPa$^{-1}$), axial conductivity spans two orders of magnitude, the three root profiles place the roots near the surface, uniformly or in two separated layers, and the rhizosphere gradient is either bypassed or resolved.

\begin{table}[p]
\centering
\caption{Model parameters.}
\label{tab:params}
\vspace{4pt}
\footnotesize
\setlength{\tabcolsep}{4pt}
\renewcommand{\arraystretch}{1.25}
\rowcolors{2}{black!5}{white}
\begin{tabular}{@{}>{\raggedright\arraybackslash}p{2.3cm}>{\raggedright\arraybackslash}p{5.2cm}>{\raggedright\arraybackslash}p{4.6cm}>{\raggedright\arraybackslash}p{3.4cm}@{}}
\toprule
\rowcolor{white}\textbf{Symbol} & \textbf{Description} & \textbf{Value} & \textbf{Source} \\
\midrule
\rowcolor{white}\multicolumn{4}{@{}l}{\textit{Soil}} \\
$\theta_{res}$, $\theta_{sat}$ & residual and saturated water content & 0.078, 0.43~m$^3$~m$^{-3}$ & \citet{carsel1988}, loam \\
$\alpha$, $n$ & retention parameters & 3.6~m$^{-1}$, 1.56 & \citet{carsel1988} \\
$K_s$ & saturated hydraulic conductivity & $2.89\times10^{-6}$~m~s$^{-1}$ & \citet{carsel1988} \\
-- & soil depth & 2~m & \\
$\psi_{min}$ & lower bound of soil water potential & $-1.5$~MPa & \\
$\theta_0$ & initial water content & 0.172 m$^3$~m$^{-3}$ ($0.4\theta_{sat}$) & \\
\rowcolor{white}\multicolumn{4}{@{}l}{\textit{Roots}} \\
$Z_R$ & rooting depth & 1~m & \citet{siqueira2008} \\
$r_r$ & root radius & 0.1~mm & \citet{siqueira2008} \\
$\bar\lambda$ & mean root length density & 4000~m~m$^{-3}$ & \citet{siqueira2008} \\
$K_r$ & root membrane permeability & $10^{-8}$~s$^{-1}$ ($10^{-9}$, $10^{-10}$~s$^{-1}$) & \citet{siqueira2008} \\
-- & $K_r(\psi_{surf})$: $P_{50}$, shape, floor (Section~\ref{sec:conductance}) & $-0.30$~MPa, 3, $10^{-3}K_r$ & \\
-- & $K_r(\psi_{surf})$ loss and recovery times & 1~h, 2~d & \\
\rowcolor{white}\multicolumn{4}{@{}l}{\textit{Xylem}} \\
$L$ & column height & 10~m & \\
$k_{max}$ & maximum xylem conductivity & $1.36\times10^{-8}$~kg~m$^{-1}$~s$^{-1}$~Pa$^{-1}$ & \citet{bohrer2005} \\
$c_1$, $c_2$ & vulnerability parameters & 4.86~MPa, 1.5 & \\
$\theta_{x,sat}$ & saturated xylem water content & 550~kg~m$^{-3}$ & \\
$P_0$, $p$ & storage parameters & 574~MPa, 20 & \citet{bohrer2005} \\
$f_C$, $f_K$ & storage and conductivity factors & 0.003--1, 0.1--10 & \\
\rowcolor{white}\multicolumn{4}{@{}l}{\textit{Leaf and forcing}} \\
LAI & leaf area index & 3 & \citet{siqueira2008} \\
$a$ & diffusivity ratio of water vapour to CO$_2$ & 1.6 & \citet{katul2010} \\
$R_d$ & dark respiration & $0.01V_{c,max}$ & \\
$V_{c,max}$ & carboxylation capacity at 25~\textdegree C & 59~\textmu mol~m$^{-2}$~s$^{-1}$ & \citet{katul2010} \\
$\lambda_1$, $\Psi_1$, $\beta$ & marginal water use efficiency, Eq.~(\ref{eq:lambda}) & 1000~\textmu mol~mol$^{-1}$, $-1$~MPa, 1.656~MPa$^{-1}$ & \\
$D$ & vapour pressure deficit & 0.010~mol~mol$^{-1}$ ($\approx1$~kPa) & \\
$c_a$ & atmospheric CO$_2$ & 350~\textmu mol~mol$^{-1}$ & \\
$Q_p$ & photon flux density & 200~\textmu mol~m$^{-2}$~s$^{-1}$ & \\
$T_L$ & leaf temperature & 20~\textdegree C & \\
$t_d$ & daylight period in Eq.~(\ref{eq:Ed}) & 12~h & \\
\bottomrule
\end{tabular}
\end{table}

\section{Results and Discussion}
The question posed in the Introduction, when a plant has water to spare for the soil, is answered here in three steps.  The first step tests the release condition of Eq.~(\ref{eq:release}) every night (Section~\ref{sec:condition}). The condition compares the soil-head contrast seen by the roots with the pull of net uptake on the collar, not that of transpiration as in \citet{couvreur2012}. The second step takes the factors named in the introduction, each read on the side of Eq.~(\ref{eq:release}) that it moves. Night-time transpiration enters through the diurnal cycle and its effect on $Q$ (Section~\ref{sec:diurnal}). The heterogeneity of soil water potential, and root pathways enters radially through the representation of the rhizosphere (Section~\ref{sec:rhizo}) and vertically through the root profile (Section~\ref{sec:profile}). The refilling of plant water storage enters through root and axial conductance (Section~\ref{sec:conductance}) and storage (Section~\ref{sec:storage}). Interactions between the different factors are also considered (ex: the representation of the rhizosphere also affects the pull by changing the water potential sensed by the roots). An ensemble run that changes one control at a time and tests the effect across the factorial is conducted. The third step attributes release to the controls and compares its size with literature reported HR (Section~\ref{sec:shapley}).

\subsection{Release condition}
\label{sec:condition}

In Eq.~(\ref{eq:release}), release begins in the driest rooted layer whenever the largest sensed soil-head contrast, $S=H-\min_k h_k$, exceeds the collar depression, termed the pull, $P=Q/A_r$. Because $Q$ carries both transpiration and xylem storage refilling (Section~\ref{sec:xylem}), daytime transpiration keeps $P$ above $S$ and every rooted layer supplies water to the xylem. 

Through most of the day, transpiration keeps $P$ above $S$, and every rooted layer supplies water to the xylem.
After dusk, $Q$ falls as the xylem store refills, and $P$ decays with it on the time scale $\tau=C/A_s$.
The timing of transpiration, the storage capacity and the axial conductance therefore act directly on $P$.
The root profile shape acts directly on $S$.
The rhizosphere gradient and the root conductance act on both (Sections~\ref{sec:rhizo} and \ref{sec:conductance}).

\begin{figure}[!htbp]
\centering
\includegraphics[width=\textwidth]{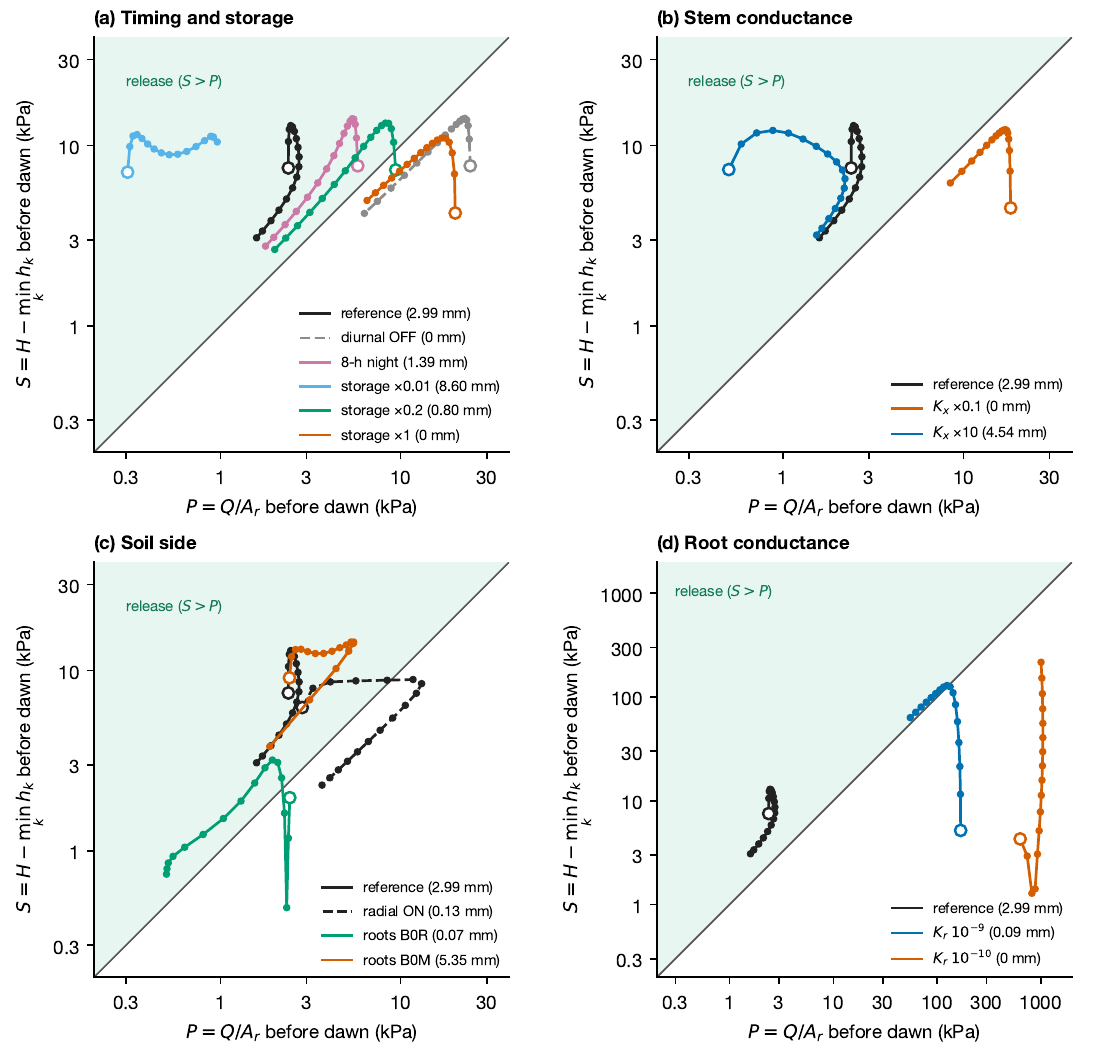}
\caption{The release condition, Eq.~(\ref{eq:release}), one hour before dawn: contrast $S=H-\min_kh_k$ against pull $P=Q/A_r$, one point for each of nights 3--19 (open circle: night 3). Black: reference (B0, $f_C=0.1$, $f_K=1$, $K_r=10^{-8}$~s$^{-1}$, 12-h night, radial OFF, diurnal ON). Other runs change one control: (a) timing of transpiration and storage, (b) axial conductance, (c) rhizosphere gradient and root profile, (d) root conductance. Above the 1:1 line at least one layer receives water. Legends: release over days 3--20.}
\label{fig:condition}
\end{figure}

In the reference run the pull fell from 36.6~kPa at dusk to 2.5~kPa one hour before dawn, while the contrast was 7.6~kPa before dawn (medians over nights 3--19).
Release began at a median of 2.9~h after dusk, once the decaying pull fell below the contrast, and it lasted until shortly after dawn.
In the reference and twelve runs that each changed one of its controls, a night released water if and only if $S>P$, on all 221 nights (Fig.~\ref{fig:condition}).
At dusk the pull exceeded the contrast on every one of these nights, by at least 1.5~kPa.

\subsection{Diurnal cycle}
\label{sec:diurnal}

Delivering the same daily transpiration at a constant rate removed release (Fig.~\ref{fig:condition}a).
The reference released 2.99~mm over days 3--20, or 3.4\% of its transpiration.  Its diurnal-OFF partner released none.  The transpiration of the two runs differed by 1.6\% (86.9 and 85.6~mm).
Across the 900 runs of the factorial (Section~\ref{sec:diag}), no diurnal-OFF run released water after the start-up transient.

Eq.~(\ref{eq:release}) explains the release condition.
With transpiration held constant, $Q$ stayed close to $T$ through the night.  The pull of the diurnal-OFF partner was 19.8~kPa at dusk and 19.1~kPa before dawn (medians), so the collar potential stayed as far below the soil potential at night as by day.
A layer stops supplying water as its head approaches the collar potential (Eq.~\ref{eq:qr}), and vertical soil flow moves water toward the lowest head.
Uptake at a steady rate, therefore, cannot dry any layer below the collar, and with no other source of contrast $S$ cannot exceed $P$.  In the diurnal-OFF partner the predawn margin $S-P$ rose from $-16.6$~kPa on night 3 to $-2.1$~kPa on night 19 without crossing zero.
With diurnal forcing, transpiration nearly stops at dusk, and the pull decays below the contrast within hours as the store refills.
Release is thus set by the decay of $Q$ after transpiration stops, and not by the daily total of transpiration.  Constant transpiration is a limit in which night-time transpiration is artificially set to daytime transpiration while preserving the daily transpiration total, and it holds the collar down in the same way.  Field and model studies agree in direction. Suppressing night-time transpiration raised HR in two of three species \citep{howard2009}.
Reverse root flow in \textit{Fraxinus} declined as night-time evaporative demand rose \citep{hultine2003}.
Nocturnal transpiration cut modelled root leakage by 25--36\% in loblolly pine plantations \citep{domec2012}.
In the terms of Eq.~(\ref{eq:release}), each of these is a larger pull before dawn.

\subsection{Rhizosphere gradient}
\label{sec:rhizo}

\begin{figure}[!htbp]
\centering
\includegraphics[width=0.6\textwidth]{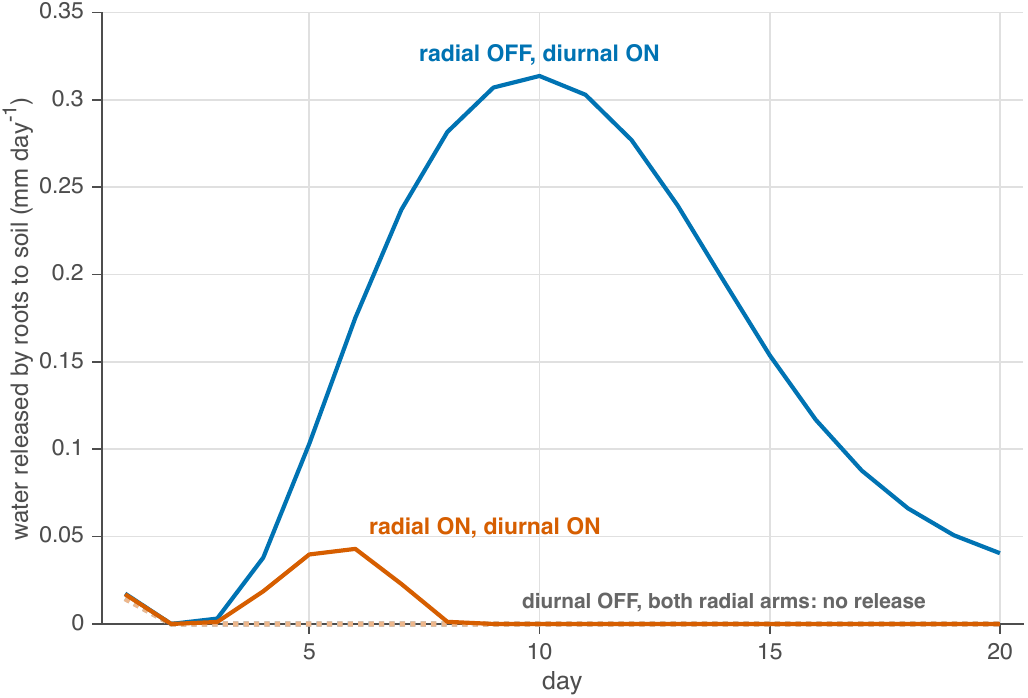}
\caption{Daily root release, B0, storage $\times0.1$, with the rhizosphere gradient bypassed (radial OFF) or resolved (radial ON) and transpiration over a 12-h day (diurnal ON) or at a constant rate with the same daily total (diurnal OFF). Over days 3--20, diurnal ON released 2.99~mm in total (3.4\% of transpiration) with radial OFF and 0.127~mm in total with radial ON. Diurnal OFF released nothing after day 2. Diurnal variations enabled HR to become a non-trivial fraction (e.g. of order 10 $\%$ of transpiration.}
\label{fig:diurnal}
\end{figure}

Resolving the rhizosphere radial gradient cut release from 2.99 to 0.127~mm at $f_C=0.1$ (96\%, Fig.~\ref{fig:diurnal}).  Across the 74 cells of the factorial in which the radial-OFF run released more than 0.01~mm, resolving the gradient cut release by 83--100\% (median 99.8\%).

\begin{figure}[!htbp]
\centering
\includegraphics[width=\textwidth]{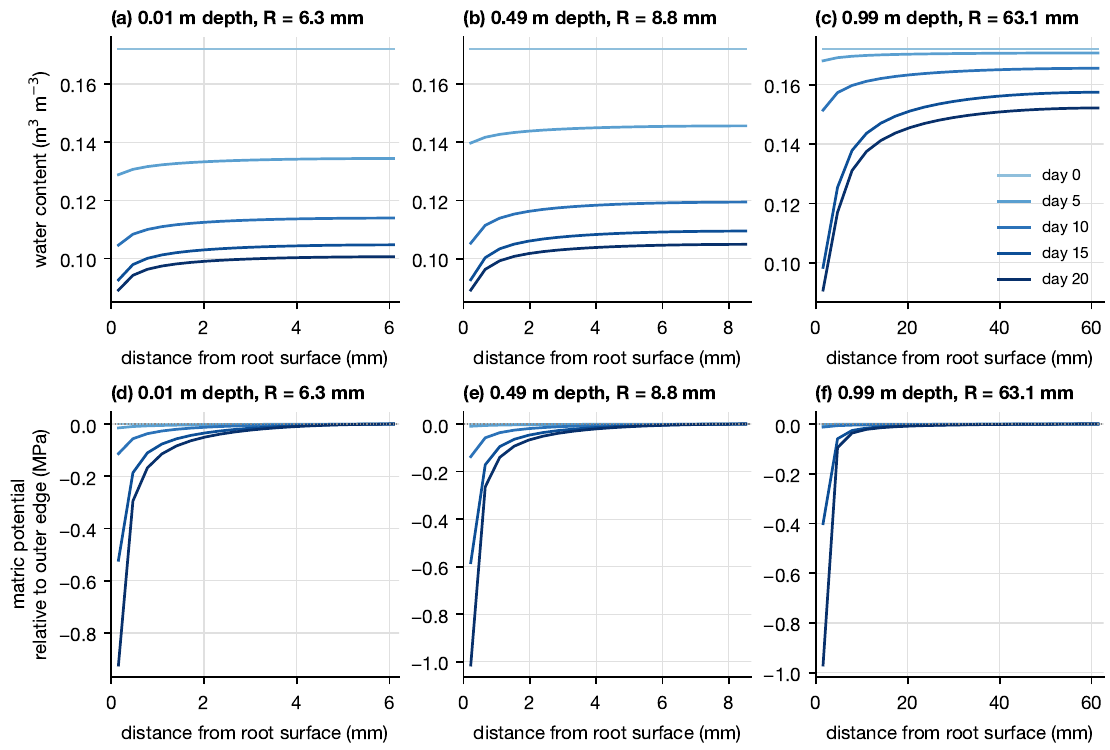}
\caption{Water content (top) and matric potential relative to the outer edge of the soil cylinder (bottom) against distance from the root surface, at 0.01, 0.49 and 0.99~m depth, 12:00 on days 5, 10, 15 and 20 (day 0: initial state). B0, storage $\times0.1$, radial ON, diurnal ON. On day 20 the drop from edge to root surface is $-0.92$, $-1.01$ and $-0.97$~MPa.  The radial dynamics reveal the interplay between water availability  and access. }
\label{fig:rhizoprofile}
\end{figure}

The gradient acts on both sides of Eq.~(\ref{eq:release}).  On the left side, the drop of water potential from the bulk soil to the root surface grew to about 1~MPa at 0.01, 0.49 and 0.99~m depth by noon on day 20 (Fig.~\ref{fig:rhizoprofile}) with part of the drop persisting through the night.
Every root surface was drawn toward the collar, so the contrast detected by the roots was smaller than with the gradient bypassed, 8.7 against 12.3~kPa one hour before dawn (medians over nights 3--7, the nights with release, Fig.~\ref{fig:condition}c).

\begin{figure}[!htbp]
\centering
\includegraphics[width=\textwidth]{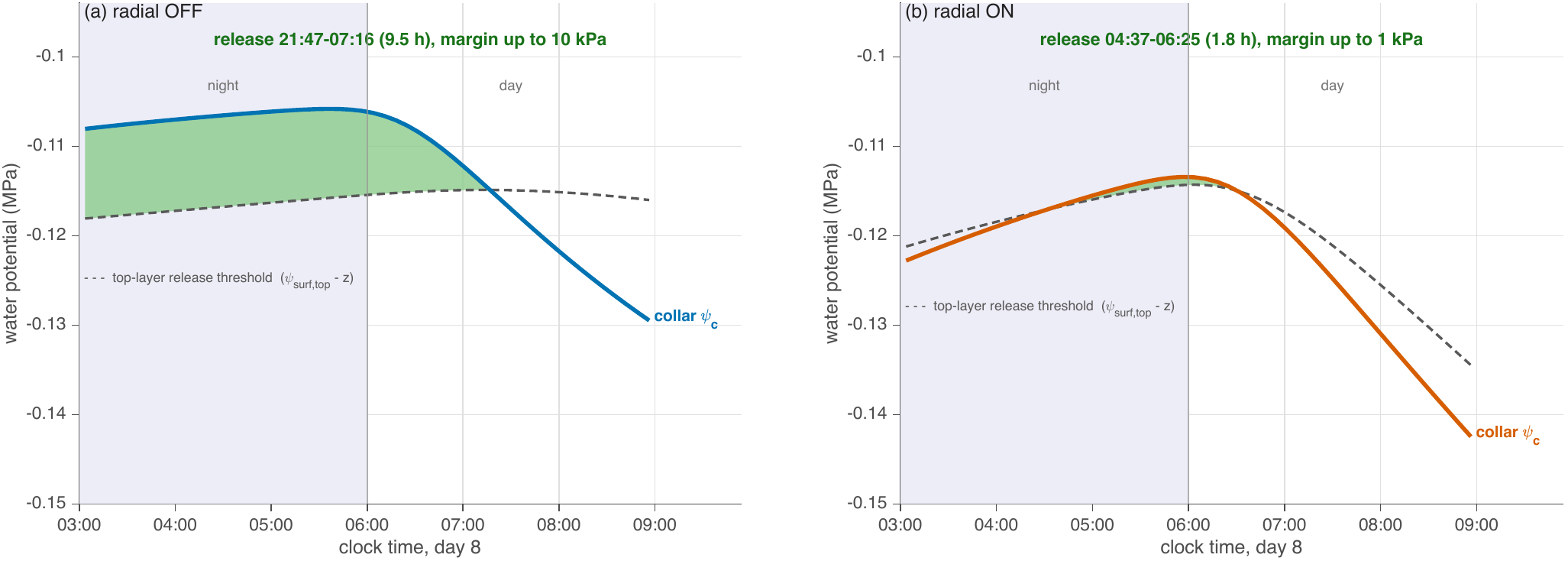}
\caption{Collar potential (solid) and top-layer release threshold (dashed), 03:00--09:00 on day 8, B0, storage $\times0.1$, diurnal ON, radial OFF (a) and ON (b). Green: release from the top layer, 21:47--07:16 (9.5~h) with a margin up to 10~kPa (OFF) and 04:37--06:25 (1.8~h) with a margin up to 1~kPa (ON).  This figure shows the direct effect on HR of rhizosphere water stress relative to bulk soil (radial-averaged) water availability.}
\label{fig:rhizo}
\end{figure}

On the right side, the rhizosphere, by accounting for the moisture heterogeneity around each root, slows the refill because it adds a resistance term in series with the root membrane.  This resistance increased as the soil dried radially around the rootlets.
Averaged over each night, the conductance of the rhizosphere and the root membrane in series fell from 0.87$A_r$ on night 3 to 0.004$A_r$ on night 19 (B0, $f_C=0.1$).
Thus, the pull before dawn rose from 2.8~kPa on night 3 to 13.1~kPa on night 9, whereas it stayed between 1.6 and 2.7~kPa with the gradient bypassed (Fig.~\ref{fig:condition}c).
The effect of that can be seen on day 8, where the smaller contrast and the larger pull together shortened release at the top layer to 1.8~h with a margin of at most 1~kPa, against 9.5~h and 10~kPa with the gradient bypassed (Fig.~\ref{fig:rhizo}).
The last release came on night 7, at a relaxation time of 7.8~h, and none came on night 8, at 11.5~h.
The B0M profile behaved the same way, with its last release on night 6, at 7.5~h.
The rhizosphere therefore ends release in two ways, by hiding the contrast from the roots and by slowing the refill so that the pull persists through the night.

\citet{siqueira2008} resolved the radial scale but did not switch it off within their model, so the two settings here give the missing comparison.
Radial OFF is the limit of a rhizosphere as wet as the bulk soil, and it therefore likely overstates HR.
In a bucket model of wild olive, rhizosphere water content drove HR, which reached 80--85\% of tree transpiration in the driest summers \citep{montaldo2022}.
There the rhizosphere is a reservoir that holds lifted water next to the roots, whereas here it is both a reservoir and a resistance.

\subsection{Root profile shape}
\label{sec:profile}

Because each run started from a uniform soil, every contrast in soil head was built by root uptake.
With linearly declining roots (B0), the dense upper layers dried first as expected. The top layer was then the first to receive water, on every night.
Uniform roots (B0R) dried all layers at similar rates.
Their contrast before dawn was 1.5~kPa, against 7.6~kPa for B0 and 12.8~kPa for the bimodal profile (B0M, Fig.~\ref{fig:condition}c).
Release followed the contrast, with 0.072~mm for B0R, 2.99~mm for B0 and 5.35~mm for B0M at $f_C=0.1$ (radial OFF).
B0M released 2.78~mm into the upper 0.3~m and 2.55~mm into the lowest 0.3~m of the root zone, around its two root modes, and 0.02~mm into the layers between them.

\begin{figure}[!htbp]
\centering
\includegraphics[width=\textwidth]{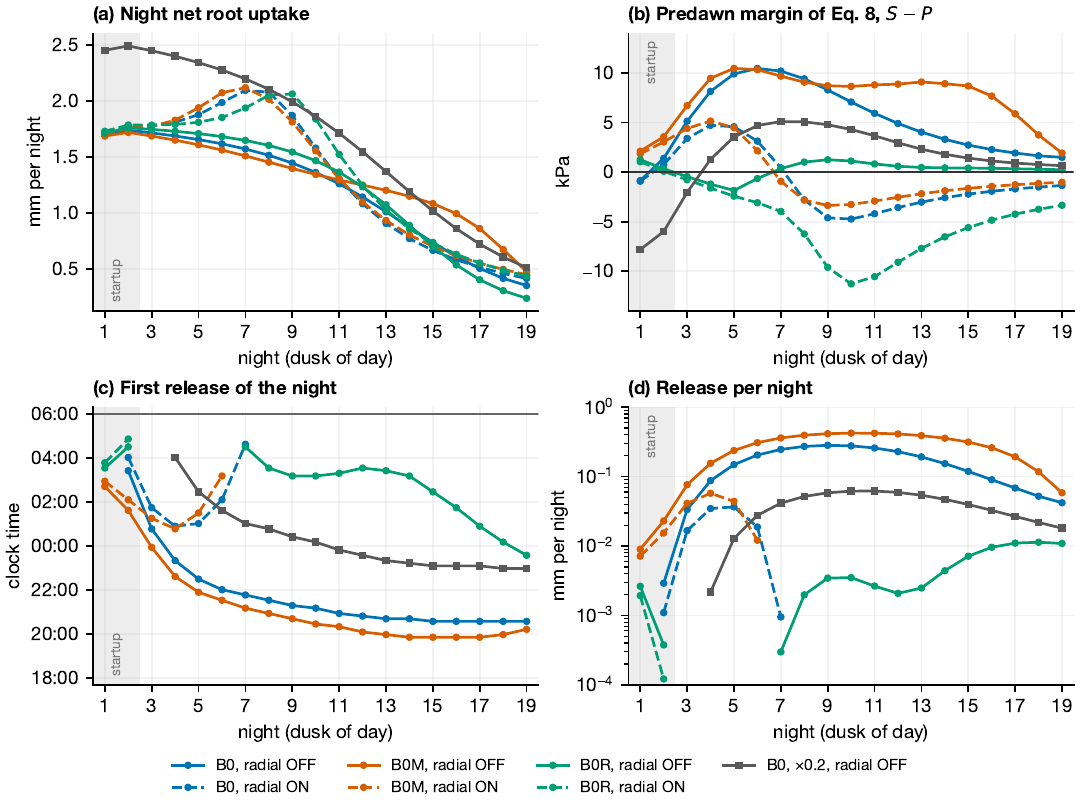}
\caption{Night-by-night course for root profiles B0 (linear), B0M (bimodal) and B0R (uniform), storage $\times0.1$ unless marked, diurnal ON, radial OFF (solid) or ON (dashed). Nights are 18:00--06:00. (a) Net root uptake. (b) Predawn margin $S-P$ (Eq.~\ref{eq:release}). (c) Clock time of the first release. (d) Release (log scale). Grey: start-up nights, the only nights on which B0R radial ON releases.  This figure highlights the significance of root profile shape on the timing and magnitude of HR. When root profile is the only cause of soil moisture non-uniformity in the vertical, HR becomes weak to non-existent for uniformly distributed roots.}
\label{fig:onset_nights}
\end{figure}

B0R released only after the pull had decayed to about 1.5~kPa, a median of 9.2~h after dusk.  It released only on nights 7--19, with predawn margins of 0.2--1.2~kPa (Fig.~\ref{fig:onset_nights}).
B0 and B0M released on 17 nights.
Their first release moved from 00:47 (B0) and 23:56 (B0M) on night 3 to 20:35 and 20:13 on night 19 as the pull at dusk declined.
Over the same nights, the night uptake that refills the xylem store fell from 1.7 to 0.35--0.48~mm.
B0M kept a predawn margin of 8.6--10.5~kPa from night 4 to night 15, and its release peaked at 0.42~mm on night 10.
Across the factorial, replacing B0 by B0R cut release by 84--100\% (median 98.1\%) in the 39 cells in which B0 released more than 0.01~mm.
Replacing B0 by B0M raised release in all 39 of these cells (median 58\%), and 10 further cells released only with B0M.

These results affirm that a concentrated root profile raises HR in a porous-media soil--plant model \citep{huang2017}. In savanna trees, predawn root water potential was below the soil water potential early in the dry season and above it later, a transition attributed to HR from deeper layers \citep{domec2006}.
The mechanism here is the weighting of $H$ in Eq.~(\ref{eq:collar}).
Roots spread evenly draw the soil down evenly and build little contrast.
Roots concentrated in separate layers dry those layers, leave a wetter layer between them, and supply the dry layers at night through the roots in the wet layer.
The predawn margin changed sign from negative to positive during the drydown, as the contrast grew while the pull before dawn declined, for B0R on night 7 and for B0 at $f_C=0.2$ on night 4 (Fig.~\ref{fig:onset_nights}b).

\subsection{Root and axial conductances}
\label{sec:conductance}

\begin{figure}[!htbp]
\centering
\includegraphics[width=\textwidth]{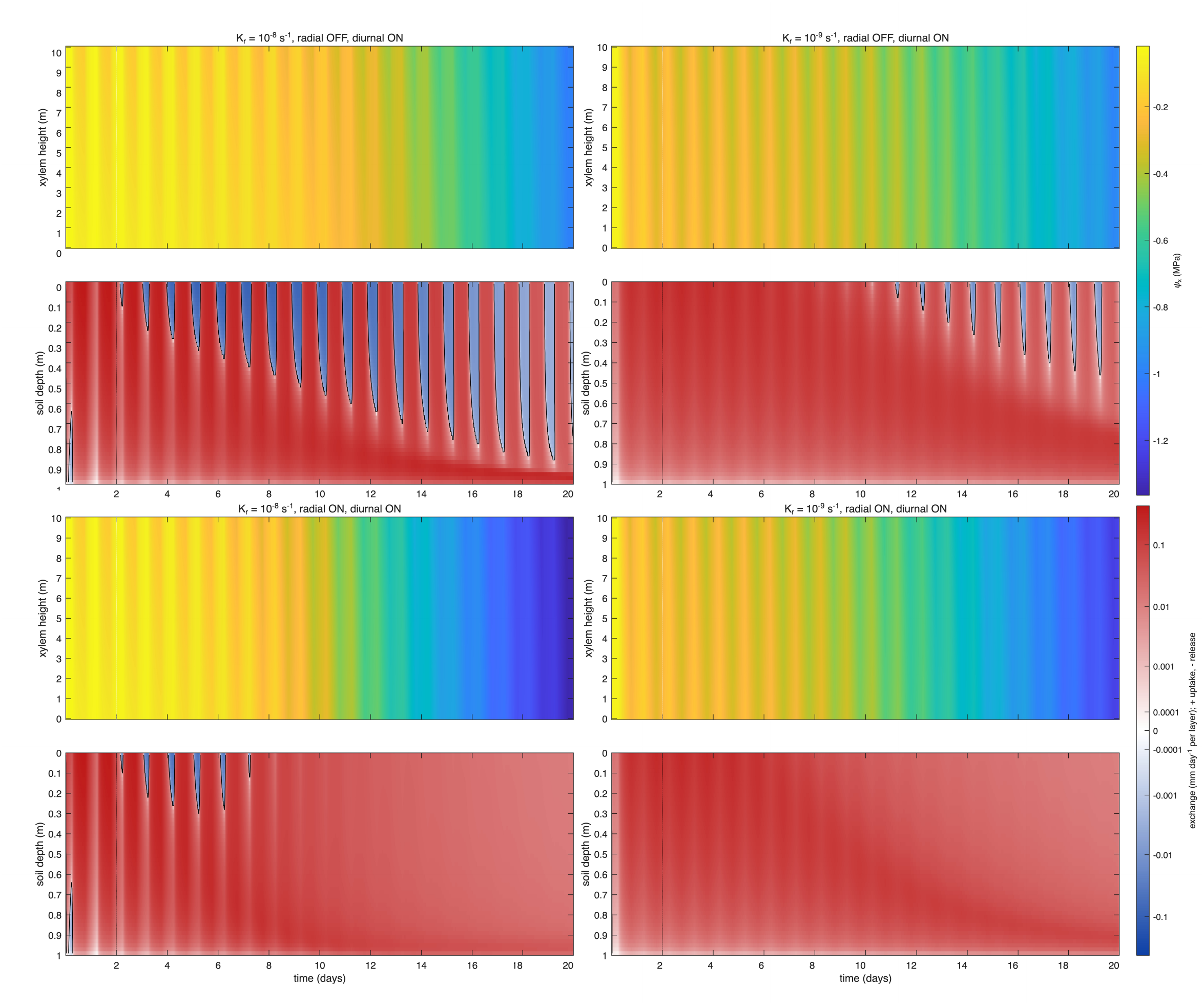}
\caption{Xylem water potential along the 10-m column (upper of each pair) and root--soil exchange by depth (lower, blue: release), B0, storage $\times0.1$, diurnal ON, $K_r=10^{-8}$ (left) and $10^{-9}$~s$^{-1}$ (right), radial OFF (top) and ON (bottom). Release over days 3--20: 2.99 and 0.091~mm with radial OFF, 0.127 and 0~mm with radial ON.  This figure shows that decreasing root conductance slows the refilling of the xylem leading to delays in the onset of HR.}
\label{fig:krlevels}
\end{figure}

Root conductance sets $A_r=K_r\sum_kw_k$, which is 221.4, 22.1 and 2.2~mm~d$^{-1}$~MPa$^{-1}$ at $K_r=10^{-8}$, $10^{-9}$ and $10^{-10}$~s$^{-1}$.
For the same uptake, a tenfold smaller $A_r$ gives a tenfold larger pull.
At $f_C=0.1$ (B0, radial OFF), the pull at dusk was 36.6, 222 and 1070~kPa at the three levels.
Release over days 3--20 was 2.99, 0.091 and 0~mm.
At $K_r=10^{-9}$~s$^{-1}$, release began only on night 11, reached no deeper than 0.45~m, and vanished when the rhizosphere gradient was resolved (Fig.~\ref{fig:krlevels}).
Across the factorial, lowering $K_r$ from $10^{-8}$ to $10^{-9}$~s$^{-1}$ cut release by 39--100\% (median 99.7\%) in the 71 cells that released more than 0.01~mm.
At $K_r=10^{-10}$~s$^{-1}$, release exceeded 0.01~mm in only 4 of 150 cells and never reached 0.04~mm.

\begin{figure}[!htbp]
\centering
\includegraphics[width=0.8\textwidth]{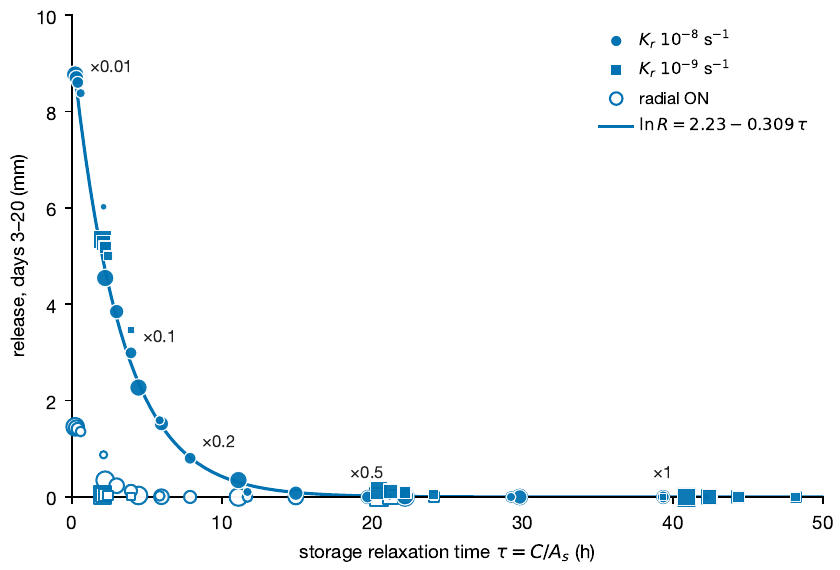}
\caption{Release over days 3--20 against $\tau=C/A_s$ for 100 B0 runs, diurnal ON, 12-h night. Circles: $K_r=10^{-8}$~s$^{-1}$. Squares: $10^{-9}$~s$^{-1}$. Labels: $f_C$ (at $K_r=10^{-8}$~s$^{-1}$, $f_K=1$, radial OFF). Marker size: $f_K$, 0.1--10. Filled: radial OFF. Open: radial ON, plotted at the $\tau$ of its radial-OFF partner. $\tau$ uses $C=18.7f_C/0.1$~mm~MPa$^{-1}$, the days 3--20 mean of the reference run scaled by $f_C$. Line: fit of $\ln R$ to $\tau$ over the 16 runs at $K_r=10^{-8}$~s$^{-1}$, radial OFF, releasing more than 0.01~mm. Not shown: 28 runs with $\tau>50$~h ($\le0.0022$~mm) and the 50 runs at $K_r=10^{-10}$~s$^{-1}$ ($\le0.0006$~mm).  This figure shows that larger xylem capacitance requires more refilling time $\tau$ which suppresses HR.}
\label{fig:taurelease}
\end{figure}

A smaller root conductance also enlarged the contrast.
With the collar far below every layer, uptake from each layer follows mainly its root density (Eq.~\ref{eq:qr}), and the dense upper layers dried without the feedback that a collar close to the soil heads imposes.
The soil contrast grew, but it was overtaken by a pull that started larger at dusk and decayed more slowly through the night (Fig.~\ref{fig:condition}d).
The contrast overtook the pull only on nights 11--19, and release was confined to those nights.
Lower root conductance also reduced lift in the storage-free model of \citet{siqueira2008}, from 1.14 to 0.80~mm~d$^{-1}$.

A root conductance that declines with the water potential at the root surface (Table~\ref{tab:params}) was run outside the factorial, at $f_C=0.1$.
It behaved as expected from Eq.~(\ref{eq:release}) and from $\tau=C/A_s$.
A conductance that falls as the soil dries lengthens the refill of the stem and narrows the path into the dry layers that receive water.
Release ended on day 15 instead of day 20, and it fell from 2.99 to 2.05~mm (radial OFF).
This agrees with the late-season decline of HR as the modelled conductance of the root pathway fell with root cavitation \citep{warren2007}.

Axial conductance enters only through $A_s$, in series with $A_r$.
At the reference, $A_r=221.4$ and $2K_x=235$~mm~d$^{-1}$~MPa$^{-1}$, so the roots and the column limit the refill about equally.
At $f_C=0.1$, release over days 3--20 was 0, 0.51, 1.50, 2.99, 4.14 and 4.54~mm for $f_K=0.1$, 0.3, 0.48, 1, 3 and 10.
Over this range of $f_K$, $\tau$ fell from 21.1 to 2.2~h (marker size in Fig.~\ref{fig:taurelease}).
At $f_K=0.1$ the pull before dawn was 16.3~kPa, above the 10.8~kPa contrast (Fig.~\ref{fig:condition}b).
Across the factorial at $K_r=10^{-8}$~s$^{-1}$, raising $f_K$ from 1 to 10 increased release by a median of 43\% in the 14 cells that released more than 0.01~mm, and 7 further cells released only at $f_K=10$.
At $K_r=10^{-9}$~s$^{-1}$, where $A_r=22.1$ is an order of magnitude below $2K_x$, the same change increased release by a median of 19\%.
Root conductance thus sets the size of the pull through $A_r$, while root and axial conductance together set the rate of its decay through $A_s$.

\subsection{Storage capacity}
\label{sec:storage}

\begin{figure}[!htbp]
\centering
\includegraphics[width=\textwidth]{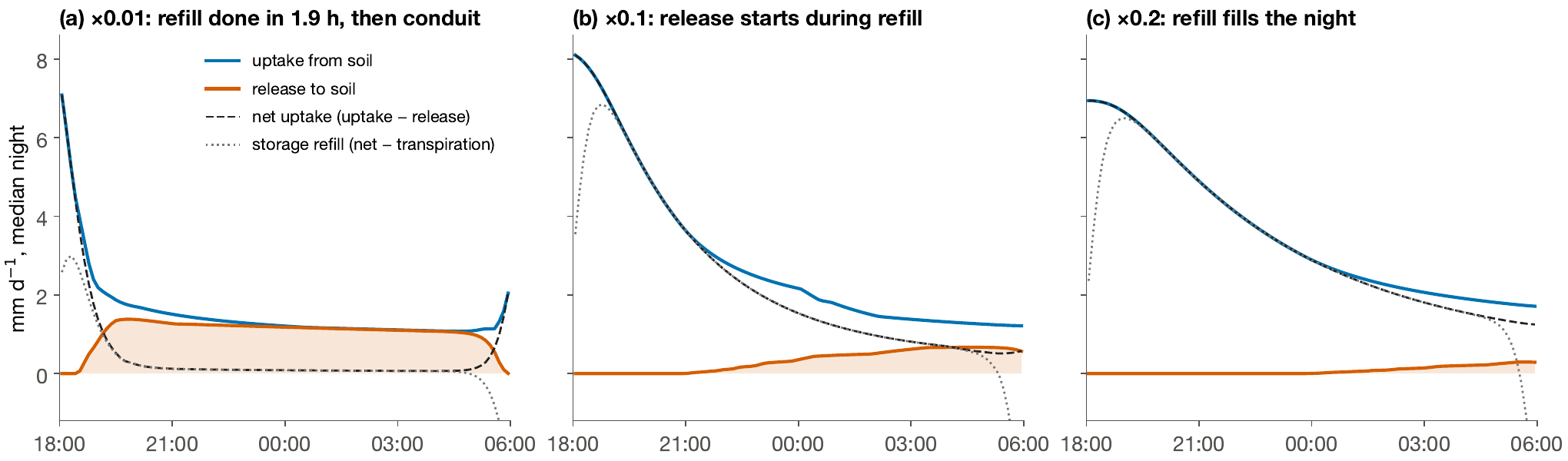}
\caption{Median night course (nights 3--19) of uptake from soil, release to soil, net uptake and storage refill, B0, radial OFF, storage $\times0.01$, $\times0.1$ and $\times0.2$. At $\times0.1$ and $\times0.2$, 64\% and 94\% of release falls before the refill ends.  Note that uptake and release can co-exist during the evening.  Water uptake from the soil directed to xylem refilling competes with HR. }
\label{fig:course}
\end{figure}

Storage capacity sets the volume that must be refilled each night through $A_s$.
At $f_C=1$ ($\tau=39.3$~h) the store took up a median of 1.63~mm per night.
No night released water.
Reducing the capacity shortened the refill and freed the night for release.
At $f_C=0.1$, 0.01 and 0.003 ($\tau=3.9$~h, 24~min and 7.1~min), the median nightly release was 0.15, 0.51 and 0.55~mm.
Its median share of that day's transpiration was 3.3, 11.0 and 13.3\%.
At $f_C=0.01$ the refill ended 1.9~h after dusk (Fig.~\ref{fig:course}a).
The roots then took up water in the wet layers and released 84--93\% of it into the dry layers, acting as a conduit (Fig.~\ref{fig:course}a).
At $f_C=0.1$ and 0.2, release began while the store was still refilling, and 64\% and 94\% of it occurred before the refill ended (Fig.~\ref{fig:course}b, c).
Release therefore does not wait for the store to be full, and refill and release can overlap.

The release elasticity to capacity, $\mathrm{d}\ln R/\mathrm{d}\ln C$, steepened from $-0.08$ between $f_C=0.003$ and 0.01 to $-0.46$, $-1.90$ and $-3.76$ between successive rungs up to $f_C=0.3$.
Release stopped between $f_C=0.3$ and 0.5.
Across the factorial, raising $f_C$ from 0.1 to 1 removed 98--100\% of release in all 28 cells that released more than 0.01~mm at $f_C=0.1$ (Fig.~\ref{fig:taurelease} for B0).
The same storage buffered the plant by day.
From $f_C=0.003$ to 1, transpiration over days 3--20 rose from 77.1 to 96.5~mm, and the lowest leaf water potential rose from $-2.55$ to $-0.31$~MPa.
Over the same range, the predawn pull rose from 0.5 to 14.0~kPa.
Storage that keeps leaf water potential high by day is the same storage that keeps the collar below the soil at night.

These results agree with the competition between storage refill and HR described in field and model studies.
A larger stem capacitance reduced HR by more than 22\% in a porous-media model \citep{huang2017}.
A single fitted whole-plant capacitance limited HR to major rainfall events \citep{vogel2017}.
A larger capacitance raised night-time root water uptake in a soil--plant model of beech and spruce \citep{martinetti2026}, as the larger refill did here (Fig.~\ref{fig:course}).
Storage withdrawal and refilling were proposed to have prevented HR in \textit{Celtis} \citep{hultine2003}.
HR in \textit{Populus euphratica} was more strongly related to stem refilling than to night-time transpiration \citep{yu2018}.
Across six sites, the effectiveness of HR in delaying soil drying diminished with increasing sapwood area per ground area \citep{meinzer2004}, which raises both storage capacity and axial conductance.
In the model, $\tau=C/A_r+C/2K_x$, so raising $C$ and $K_x$ by the same factor lengthens $\tau$ through the root term and reduces release, in the same direction.
In savanna trees, the predawn disequilibrium between leaves and soil increased with stem capacitance toward a fitted asymptote of 0.15~MPa, and the species with the lowest capacitance equilibrated before the end of the night \citep{scholz2007}.
The model shows the same rise of the predawn pull with capacity, because a larger store is still refilling at dawn and so draws a larger $Q$. In that context, the restrictions on release and the dawn disequilibrium are the effect of the same knob.

\subsection{Attribution and comparison with reported release}
\label{sec:shapley}

\begin{figure}[!htbp]
\centering
\includegraphics[width=\textwidth]{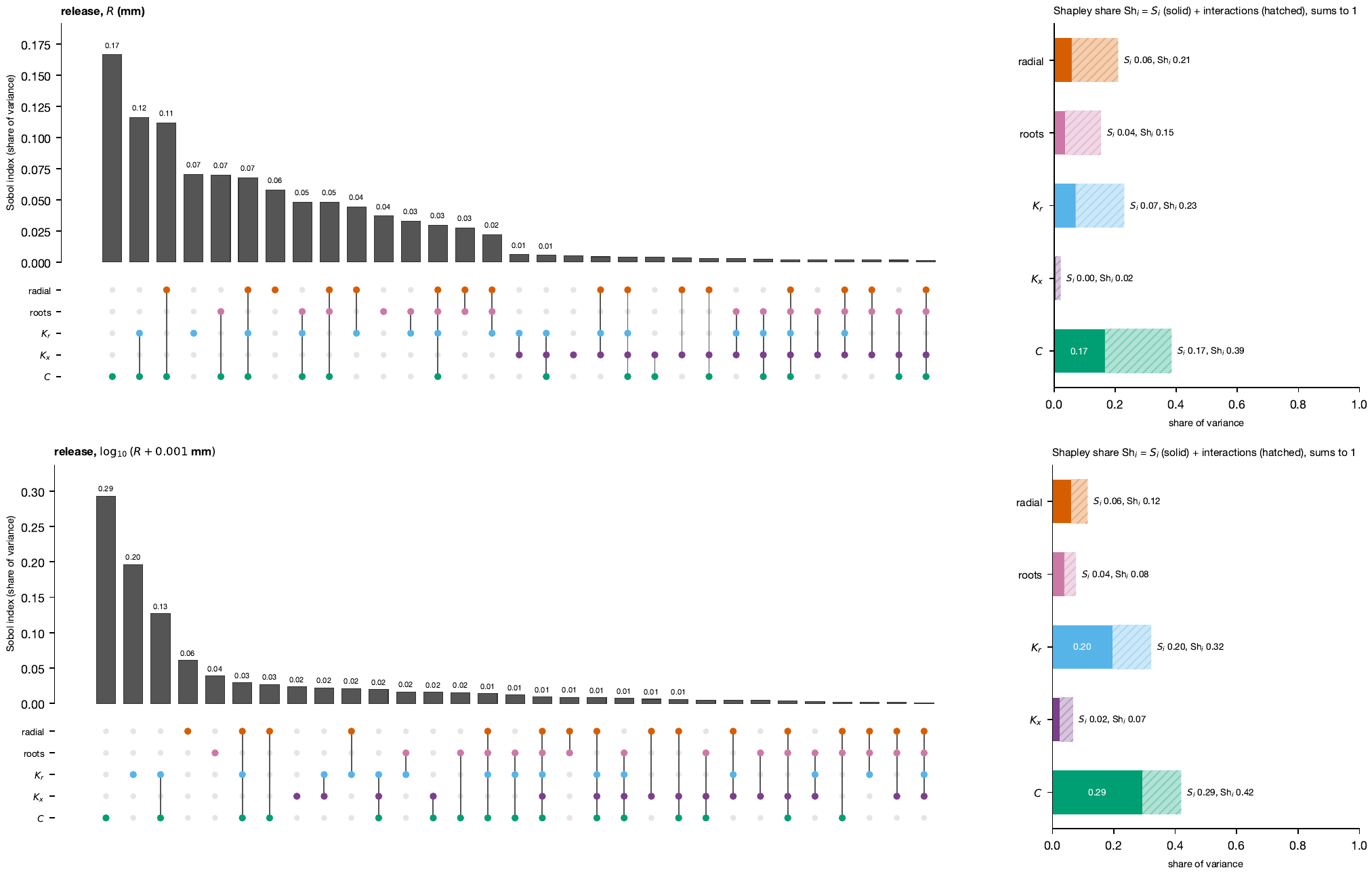}
\caption{Attribution of release over days 3--20 to five controls, 450 diurnal-ON runs, 12-h night: rhizosphere gradient (OFF, ON) $\times$ root profile (B0, B0R, B0M) $\times$ $K_r$ ($10^{-8}$, $10^{-9}$, $10^{-10}$~s$^{-1}$) $\times$ $f_K$ (0.1, 0.5, 1, 2, 10) $\times$ $f_C$ (0.01, 0.1, 0.2, 0.5, 1). Left: variance share of each of the 31 Sobol terms (dots: controls in the term), summing to 1. Right: Shapley share $\mathrm{Sh}_i$ of each control, the first-order index $S_i$ (solid) plus an equal split of each interaction term that contains the control (hatched), summing to 1 over the five controls.  The dominant process on HR is the refilling of xylem storage.  Beyond this dominant process, HR is impacted by the presence of radial dynamics and by the magnitude of root conductance.}
\label{fig:shapley}
\end{figure}

The exact decomposition of $\log_{10}(R+0.001$~mm$)$ over the 450 diurnal-ON runs assigned Shapley shares of 0.42 to storage capacity, 0.32 to root conductance, 0.12 to the rhizosphere gradient, 0.08 to the root profile and 0.07 to axial conductance (Fig.~\ref{fig:shapley}).
The first-order indices summed to 0.61, so interactions held 39\% of the variance.
The largest terms were storage alone (0.29), root conductance alone (0.20) and their interaction (0.13).
For release in millimetres the shares were 0.39, 0.23, 0.21, 0.15 and 0.02 in the same order.
Here, interactions then held 66\% of the variance.
The interaction of storage and root conductance is consistent with their joint action through $\tau=C/A_s$, because $A_s$ depends on $A_r$.
The shift between the two representations suggests that storage and root conductance mainly decide whether release occurs, which dominates the logarithm, while the rhizosphere gradient and the root profile decide the size of the largest releases, which gain weight in millimetres.
Axial conductance held the smallest share because it changes $A_s$ appreciably only where $A_r$ is large, at $K_r=10^{-8}$~s$^{-1}$.

The synthesis of 47 published estimates tested transpiration, root length, water-table depth and a humidity index as predictors \citep{yang2022}.
Those predictors co-vary among sites, whereas the controls here were crossed in a full factorial.
Storage and root conductance were not among those predictors, and here they held three quarters of the Shapley share of $\log_{10}R$.
The shares describe the tested ranges, each spanning one to two orders of magnitude, and they would change with other ranges or another representation of release \citep{owen2014}.

\begin{figure}[!htbp]
\centering
\includegraphics[width=\textwidth]{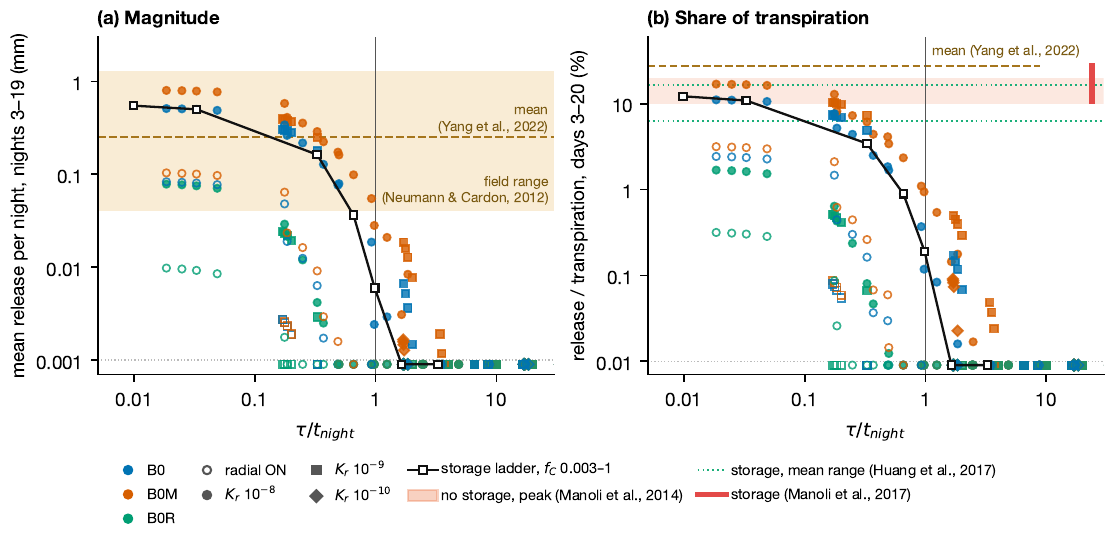}
\caption{Release of the 450 diurnal-ON factorial runs and of the reference storage ladder against $\tau$ (as in Fig.~\ref{fig:taurelease}) and reported HR. (a) Mean release per night (nights 3--19), with field magnitudes \citep{neumann2012} and the 47-site mean \citep{yang2022}. (b) Release over transpiration (days 3--20), with the peak shares of a model without storage \citep{manoli2014}, the mean shares \citep{huang2017} and range \citep{manoli2017} of models with storage, and the mean \citep{yang2022}. Colour: root profile. Symbol: $K_r$. Open: radial ON. Dotted floor: runs below 0.001~mm (a) or 0.01\% (b). Vertical line: $\tau=t_{night}$. Black line and squares: storage ladder (B0, $K_r=10^{-8}$~s$^{-1}$, $f_K=1$, radial OFF, $f_C=0.003$--1, adding $f_C=0.003$ and 0.3).  The proposed model and reported experiments agree on the magnitude of HR (panel a) when the dimensionless refilling time $\tau/t_{night}\leq1$.  Despite the limited number of plant variables analyzed here, the modeled magnitude of HR relative to daytime transpiration bounds prior experiments (panel b).}
\label{fig:literature}
\end{figure}

The model reaches the magnitudes of HR reported in the field only in part of the control space (Fig.~\ref{fig:literature}).
Mean release per night fell within the field range of 0.04--1.3~mm~d$^{-1}$ \citep{neumann2012} in 48 of the 450 diurnal-ON runs.
It exceeded the mean of 0.249~mm~d$^{-1}$ across 47 sites \citep{yang2022} in 22 runs, all with $f_C\le0.1$, the rhizosphere gradient bypassed and root profile B0 or B0M.
In the reference run, release peaked at 0.28~mm on night 9, within the peaks of 0.16--0.30~mm~d$^{-1}$ reported for old-growth conifers \citep{warren2007} and below the 1.05~mm~d$^{-1}$ peak in a loblolly pine plantation \citep{domec2010}.

Shares of transpiration were more restrictive.
Release over days 3--20 exceeded 10\% of transpiration in 12 of the 450 runs, all with $f_C=0.01$, the gradient bypassed and root profile B0 or B0M.
The largest share was 17.0\%, and none reached the mean contribution of 27.4\% \citep{yang2022}.
None of the runs with the gradient resolved exceeded 3.2\%.
Along the storage ladder of root profile B0 (black line in Fig.~\ref{fig:literature}b), release reached the 10--20\% of the storage-free model of \citet{manoli2014} only at $f_C\le0.01$ ($\tau\le24$~min), with 10.9\% at $f_C=0.01$ and 12.2\% at $f_C=0.003$.
With B0M roots, shares above 10\% occurred at $\tau$ up to 2.2~h.
The 37\% and 25\% of potential transpiration of \citet{siqueira2008} and the 19\% of \citet{ryel2002} were obtained without storage so they represent the upper limit of HR.
Release in the model is therefore smaller than the reported shares except where storage is small, the roots are contrasted and the rhizosphere offers no resistance.
Two features of the runs would make release here lower than at many field sites.
The soil started uniform, and soil texture was assumed homogeneous so the contrast was mainly created by the root density profile.

\section{Limitations and Conclusions}
\label{sec:limits}

Several limitations should be considered before drawing conclusions. The simulations used a single loam soil and a forcing regime repeated each day, with 20~d drydowns starting from a uniform soil water profile and excluding rainfall, soil evaporation, and groundwater contributions. Night-time transpiration was represented only by the tails of the prescribed diurnal function, with 2.8\% of daily transpiration occurring outside 06:00--18:00. The marginal water-use efficiency $\lambda_e$, used to calculate stomatal conductance, was computed once per day and held constant within that day. Root hydraulic conductance was held 
constant except in the $K_r(\psi_{surf})$ runs, and the rhizosphere had the properties of the bulk soil.

Within these limits, the work established a mechanistic release criterion for HR in a transpiring plant with storage. Root water release occurred when the soil-head contrast exceeded the residual pull exerted by net root uptake on the collar before dawn. Unlike previous formulations that relied on transpiration alone, the criterion explicitly accounts for xylem refilling and its storage dynamics thereby linking HR to the full soil-root-xylem system. The model demonstrated that plant water  storage both buffers xylem tension during the day and delays water release into the soil at night. Furthermore, HR computed without accounting for plant water storage must be viewed as an upper bound under the simulated conditions.

Across all 221 nights tested, roots released water to the soil whenever the soil hydraulic head contrast exceeded the opposing hydraulic head difference associated with  net root uptake. This condition was generally reached as net root uptake declined during the night, mostly when the storage relaxation time $\tau=C/A_s$ was shorter than or comparable to the length of the night.

The study introduced an exact Sobol-Shapley decomposition of HR in a fully coupled soil-root-xylem framework.  The  resulting Shapley shares quantify   the relative contributions of the controls across the tested parameter levels, with all configurations weighted equally. They therefore describe the experimental design rather than a ranking of controls across a population of plants.   Applying such Sobol-Shapley decomposition revealed that storage capacity and root conductance held three quarters of the Shapley share of $\log_{10}$ release, and a uniform root profile or a resolved rhizosphere gradient typically reduced release by more than an order of magnitude. The analysis also revealed that interactions among controls account for a large fraction of the variability in HR. Approximately 39\% of the variance arose from interactions. An immediate consequence of this outcome is that HR cannot be understood from isolated factors and must be treated as an 'emergent' property of the coupled soil, root, rhizosphere, xylem, and storage processes.

Future work seeks to expand this model in two directions. (1) developing reduced-order models, such as an equivalent resistor--capacitor circuit, to predict the onset and cumulative volume of water release during prolonged drydowns and clarify interactions among plant storage, soil--root hydraulics, and xylem hydraulics; and (2) incorporating phloem into the current xylem--leaf framework.  This addition would account for the fate of photosynthetically derived  sugars and their role in mediating hydraulic and microbial processes active in the rhizosphere. Root exudates remain a challenge to quantify experimentally and their role in modifying hydraulic redistribution remains unexplored.

\clearpage

\bibliographystyle{abbrvnat}
\bibliography{references}
\end{document}